\documentclass[10pt]{article}
\pdfoutput=1

\usepackage{mathrsfs}
\usepackage{booktabs,tabularx,array,longtable}
\usepackage{setspace}
\usepackage[letterpaper,top=25mm,left=28mm,bottom=27mm,right=28mm]{geometry}
\usepackage{titlesec}
\titleformat*{\section}{\Large\bfseries}
\titleformat*{\subsection}{\large\bfseries}
\titleformat*{\subsubsection}{\bfseries}
\titleformat*{\paragraph}{\bfseries}
\usepackage{booktabs}
\usepackage{tikz}

\definecolor{survey-1}{HTML}{3366A4}
\definecolor{survey-3}{HTML}{237A45}
\definecolor{survey-4}{HTML}{B5A31B}
\definecolor{survey-5}{HTML}{DB566D}
\definecolor{survey-6}{HTML}{9A3576}
\definecolor{survey-7}{HTML}{54368A}
\definecolor{survey-8}{HTML}{B66A22}
\definecolor{survey-9}{HTML}{8D253B}
\definecolor{survey-11}{HTML}{CB4B28}
\definecolor{survey-12}{HTML}{717E2D}
\definecolor{survey-13}{HTML}{BE589E}
\definecolor{survey-14}{HTML}{227C8D}
\definecolor{survey-15}{HTML}{5046A5}
\definecolor{survey-16}{HTML}{B18A6A}
\definecolor{survey-18}{HTML}{72933C}
\definecolor{survey-19}{HTML}{66504B}
\definecolor{survey-20}{HTML}{DA8E86}
\definecolor{survey-22}{HTML}{6C87C6}

\newcommand{\SurveyDot}[1]{%
  \tikz[baseline=-.5ex]{%
    \fill[survey-#1] (0,0) circle[radius=2.9pt];
    \ifnum#1=20
      \fill[white] (0,0) circle[radius=1.25pt];
    \fi
    \ifnum#1=22
      \fill[white] (0,0) circle[radius=1.25pt];
    \fi
    \draw[black!45,line width=.18pt]
      (0,0) circle[radius=2.9pt];
  }%
}



\usepackage{soul}

\usepackage{mathtools}
 \usepackage{tcolorbox}
    \tcbuselibrary{listings,theorems}

\usepackage{upgreek}

\usepackage{empheq}
\usepackage{microtype}
\usepackage{relsize}
\usepackage[dvipsnames]{xcolor}
\usepackage{cite}
\usepackage{enumitem}
\usepackage{ifpdf}
\usepackage{lmodern}
\usepackage[T1]{fontenc}
\usepackage[ansinew]{inputenc}
\usepackage{amsmath}
\usepackage{amsthm}
\usepackage{relsize}
\usepackage{amssymb}
\usepackage{comment}
\usepackage{tikz}
\usepackage{graphicx}
\usepackage{color}
\definecolor{darkblue}{cmyk}{0.9,0.9,0,0}
\definecolor{darkgreen}{rgb}{0,0.55,0}
\usepackage[colorlinks=true,linkcolor=darkblue,citecolor=darkblue,urlcolor=darkblue]{hyperref}
\usepackage{epsfig}
\usepackage{epstopdf}
\usepackage{graphicx}

\makeatletter
\long\def\@makecaption#1#2{
  \vskip\abovecaptionskip
  \sbox\@tempboxa{{\captionfonts #1: #2}}
  \ifdim \wd\@tempboxa >\hsize
    {\captionfonts #1: #2\par}
  \else
    \hbox to\hsize{\hfil\box\@tempboxa\hfil}
  \fi
  \vskip\belowcaptionskip}
\makeatother

\def\cref{\ref}
\setlist[itemize]{leftmargin=1.45em,itemsep=0.25em,topsep=0.25em}
\setlist[enumerate]{leftmargin=1.7em,itemsep=0.7em,topsep=0.3em}
\renewcommand{\arraystretch}{1.24}
\newcolumntype{L}[1]{>{\raggedright\arraybackslash}p{#1}}

\def\cT{\mathcal{T}}
\newcommand{\Dgap}{\Delta_{\mathrm{gap}}}

\newcommand{\cF}{\mathcal F}
\newcommand{\cW}{\mathcal W}
\newcommand{\Wp}{\mathcal W_+}
\newcommand{\sP}{\mathscr P}
\newcommand{\Lam}{\Lambda}

\newcommand{\im}{\operatorname{im}}

\newcommand{\R}{\mathbb R}
\newcommand{\Z}{\mathbb Z}
\newcommand{\N}{\mathbb N}
\newcommand{\dc}{\delta c}
\newcommand{\ads}{\mathrm{AdS}}

\newcommand{\Vol}{\operatorname{Vol}}

\renewcommand{\arraystretch}{1.12}
\setlist{nosep}
\def\mhs{M_{\rm HS}}
\def\rsa{\rightsquigarrow}

\def\bJ{{\boldsymbol J}}

\def\bitem{\begin{itemize}}
\def\eitem{\end{itemize}}
\def\benum{\begin{enumerate}}
\def\eenum{\end{enumerate}}

\definecolor{greencirc}{rgb}{0, 0.73, 0.5}

\def\enumcom{\begin{enumerate}[label*=  \sf (\roman*),wide, labelwidth=!, labelindent=0pt]}

\def\cS{\mathcal{S}}

\def\vareps{\varepsilon}

\def\cF{\mathcal{F}}

\def\half{{1\o2}}

\def\R{\mathbb{R}}

\def\x{\times}

\def\eps{\epsilon}

\def\Z{\mathbb{Z}}

\def\1{{\rm 1-loop}}

\def\c{\cite}

\def\cA{\mathcal{A}}

\def\cW{\mathcal{W}}

\def\cM{\mathcal{M}}

\def\cN{\mathcal{N}}
\def\cA{\mathcal{A}}

\def\c{\cite}

\def\vs{\vskip .1 in}

\def\o{\over}
\def\g{\gamma}
\def\D{\Delta}
\def\rar{\rightarrow}

\def\eqr{\eqref}

\def\cO{{\cal O}}

\def\i{\infty}

\def\ssec{\subsection}
\def\sssec{\subsubsection}
\def\sec{\section}

\def\foot{\footnote}
\newcommand{\es}[2] {\begin{equation} \label{#1} \begin{split} #2 \end{split} \end{equation}}
\newcommand{\e}[2] {\begin{equation} \label{#1} #2 \end{equation}}

\newcommand{\beq}{\begin{equation}}
\newcommand{\eeq}{\end{equation}}
\newcommand{\beqy} {\begin{eqnarray}}
\newcommand{\eeqy} {\end{eqnarray}}
\newcommand{\bsmat}{\begin{smallmatrix}}
\newcommand{\esmat}{\end{smallmatrix}}
\newcommand{\bmat}{\begin{matrix}}
\newcommand{\emat}{\end{matrix}}

\def\({\left(}
\def\){\right)}
\def\[{\left[}
\def\]{\right]}

\def\<{\langle}
\def\>{\rangle}

\def\a{\alpha}
\def\b{\beta}
\def\g{\gamma}

\def\d{\delta}
\def\D{\Delta}

\renewcommand{\l}{\lambda}
\def\l{\lambda}

\def\t{\tau}

\def\vs{\vskip .1 in}

\usepackage{varioref}
\usepackage{makeidx}
\makeindex

\usepackage[english]{babel}
\usepackage{caption}
\usepackage{subfig}

\numberwithin{equation}{section}

\usepackage{tabularx}
 
\begin{document}

\begin{spacing}{1.15}

\begin{titlepage}
\vspace*{2cm}
\begin{center}

{\Large \bfseries Gravity as Large Scale Structure: \\ \vspace*{3mm} Families of CFTs and the Higher-Spin Gap}

\vspace*{1cm}

Eric Perlmutter

\vspace*{2mm}

\textit{\small Institut de Physique Th\'eorique, Universit\'e Paris-Saclay, CEA, CNRS\\Orme des Merisiers, 91191, Gif-sur-Yvette Cedex, France\\
\vskip .2 in
Institut des Hautes \'Etudes Scientifiques\\ 
35 Route de Chartres, 91440, Bures-sur-Yvette, France}

\vspace*{1cm}

\end{center}

\begin{abstract}

We observe an apparently universal behavior in large $N$ conformal field theory (CFT) in $d\geq 2$ dimensions: the single-trace, higher-spin gap obeys the parametric bound $\Delta_{\rm gap} \lesssim (\delta c)^{1/d}$, where $\delta c$ is the central charge increment between successive theories. The origin of this bound is explained heuristically, using a generalized notion of branes that interpolate between conformal points in theory space. Seeking a more precise framework for this picture leads to the central concept of families of CFTs, which tacitly underlies various features and puzzles of AdS/CFT. We develop a working definition of families, with inspiration from families of $L$-functions in number theory, and formulate a family-level version of the gap bound. Interpreted holographically for families of AdS$_{d+1}$ vacua, this bound, when applicable, rules out Planckian scale separation, constrains the size and number of extra dimensions, and explains simple scaling patterns in string/M-theory. It also accounts for why all known large $N$ vector models have a bounded gap. 

\end{abstract}

\end{titlepage}
\end{spacing}

\pagenumbering{roman}
\begin{spacing}{0.6}

\setcounter{tocdepth}{2}
\tableofcontents
\end{spacing}

\newpage

\pagenumbering{arabic}
\setcounter{page}{1}

\begin{spacing}{1.07}

\sec{Introduction}

We begin this paper with an invitation to scrutinize certain longstanding mores of semiclassical AdS/CFT.

Over the last 30 years or so, many a paper on AdS/CFT includes an introductory passage that reads something like this: 
\begin{quote}
{\textit{In the AdS/CFT correspondence, in the semiclassical limit, the holographic dictionary relates Newton's constant in the bulk to the number of degrees of freedom in the CFT as $G \sim 1/N^z$, where the exponent $z$ depends on the theory.}}
\end{quote}
The authors may then proceed, in the name of concreteness, to use (say) the gauge theory scaling $z=2$. 

This passage, though familiar, obscures a core conceptual point: the relation between $G$ and $N$ is, in the first place, a {\it bulk} relation. The quantity $N$ has a bulk meaning of its own, typically as a flux or brane number, whose relation to $G$ is enforced by the gravitational field equations. The above presentation conflates two distinct entries of the dictionary: first, $G$ in the bulk maps to $1/c$, the inverse central charge of the CFT; and second, $N$ in the bulk maps to a parameter $N$ in the CFT, which in turn has a purely field-theoretic relation to $c$. That relation, and in particular the exponent $z$, is an inherent datum of the CFT in question -- suggesting that an arbitrary choice of $z$ is missing something. 

While this might seem pedantic, disentangling these concepts immediately raises the first of four fundamental questions that motivated this work:

\begin{enumerate}[label*=  \bf \arabic*)]
\item {\it What is $N$?} In examples we understand, $N$ often labels the rank of a gauge or global symmetry group. But from the abstract, non-perturbative CFT point of view, it is not obvious what, or even why, $N$ is.

\item {\it Why is $c \sim N^{d/2}$ in the most canonical examples of M2-, D3- and M5-brane holography?} Is there an underlying principle obeyed by these CFTs, or a bound that they extremize? We are not aware of a fully satisfactory explanation of this well-known pattern, in which both the CFT meaning of $N$, and the bulk theory, differ for each example.

\item {\it Why are known large $N$ vector models, defined here as CFTs for which $z=1$, never strongly coupled?} We observe that vector models have either a slightly broken higher-spin symmetry, like the critical $O(N)$ model in $d=3$, or, at most, an intermediate value of the effective coupling.

\item {\it When is a CFT at large-but-finite $c$ well-approximated by its $1/c$ expansion, i.e. by a semiclassical theory of AdS gravity?}

\end{enumerate}

All of these questions hover around a concept that has been mostly implicit in AdS/CFT literature: that of {\it families} of CFTs, and of AdS vacua of quantum gravity. By definition, semiclassical gravity is defined by a parametric limit $G \rar 0$; so any question about the semiclassical gravitational path integral really needs to be understood in the sense of a sequence of theories, and likewise on the CFT side as $c\rar\i$. This has been reinforced by recent, more sophisticated investigations in the literature on closed universes in AdS/CFT \c{ASS,UsatyukZhaoClosed, AntoniniRathUniqueness,EngelhardtGesteauHarlowObserver, AntoniniRathSasietaSwingleVilar,EngelhardtGesteauFurtherEvidence, LiuClosedUniverses, GesteauNoGoClosed,BelinDeBoerBabyUniverses,KFWpre}, and on the apparent averaging that semiclassical gravitational wormhole amplitudes seem to require \c{Maldacena:2004rf, Saad:2018bqo, SSS, MM, CJ, CCHM, SW, LiuFiltering}. These puzzles, we submit, have arisen from our ignorance of the family aspect of holography. One active research question is whether the non-factorizing multi-boundary amplitudes of gravity are signatures of, or imply a version of, erraticness of CFT data on exponentially fine scales \c{LiuFiltering, KFW, MWang, 2607, Liu:2026fnd, HiranoThirdQuantized, KlingerFactorization}; in this picture, the putative erraticness seems to be {\it in $N$}, and should not be thought of as belonging to a ``single theory'', which has fixed central charge. Indeed, we propose that these types of questions, which are probing notions of correlation in theory space itself, should properly be viewed conceptually as {\it family-level} questions, at asymptotically large central charge.

The concept of families is also directly relevant for Question 4 above, in the following sense. If a definition of family endows a sequence of theories with a certain structure, then it may be capable of differentiating among those large-but-finite $c$ theories for which the $1/c$ expansion is a good approximation, and those for which it is not. 

In fact, one need not look far for examples of the latter. Consider the D1--D5 CFT, which will turn out to be an instructive setting for our later computations. This theory is characterized by two positive integers $(N_1,N_5)$, with central charge $c = 6N_1N_5$ up to an $N_i$-independent integer shift (irrelevant for our purposes). The moduli space $\cM_{\text{D1--D5}}$ of this theory is intricate, governed by U-duality symmetries. It is long-known \c{Larsen:1999uk}, if underappreciated, that its structure is sensitive to {\it number-theoretic} properties of $c/6$: the set of strongly coupled cusps of $\cM_{\text{D1--D5}}$ is determined by factorizations of $c/6$ into coprime pairs $(N_1,N_5)$. In particular, basic entries of the D1--D5 holographic dictionary (cf. \eqr{eq:D1D5-RR-frame}) imply that $\cM_{\text{D1--D5}}$ does not possess a supergravity point if $c/6$ is not factorizable into two parametrically large integers $(N_1,N_5)$. Therefore, in, say, the case of $c/6$ prime and large -- indeed, {\it everywhere} along an unbounded arithmetic sequence of prime $c/6$ -- the bulk dual looks nothing like semiclassical supergravity! Exposing these kinds of distinctions, and understanding their physical significance in AdS/CFT, is one aim of our work.\foot{In our estimation (and at serious risk of further pedantry), it is an unfortunate historical fact, inherited from the presence of a symmetric orbifold point on $\cM_{\text{D1--D5}}$, that the product $N_1N_5$ is commonly abbreviated simply as $N$. This has, perhaps, led to an obfuscation of some of the properties of the D1--D5 CFT that we discuss in this paper.}  

In this paper, we parlay these abstract considerations into a concrete, empirical observation about the spectrum of known large $N$ CFTs (holographic and otherwise). From a pure CFT point of view, the result at first seems surprisingly strong. We explain why this observation holds, and explore the conditions under which this ought to be true more generally. In doing so, we initiate an exploration of families of CFTs, and formulate a family-level extension of the aforementioned bound. When applicable, its implications for holography are significant. 

\begin{figure}[!htbp]
\centering
{\includegraphics[scale=.85]{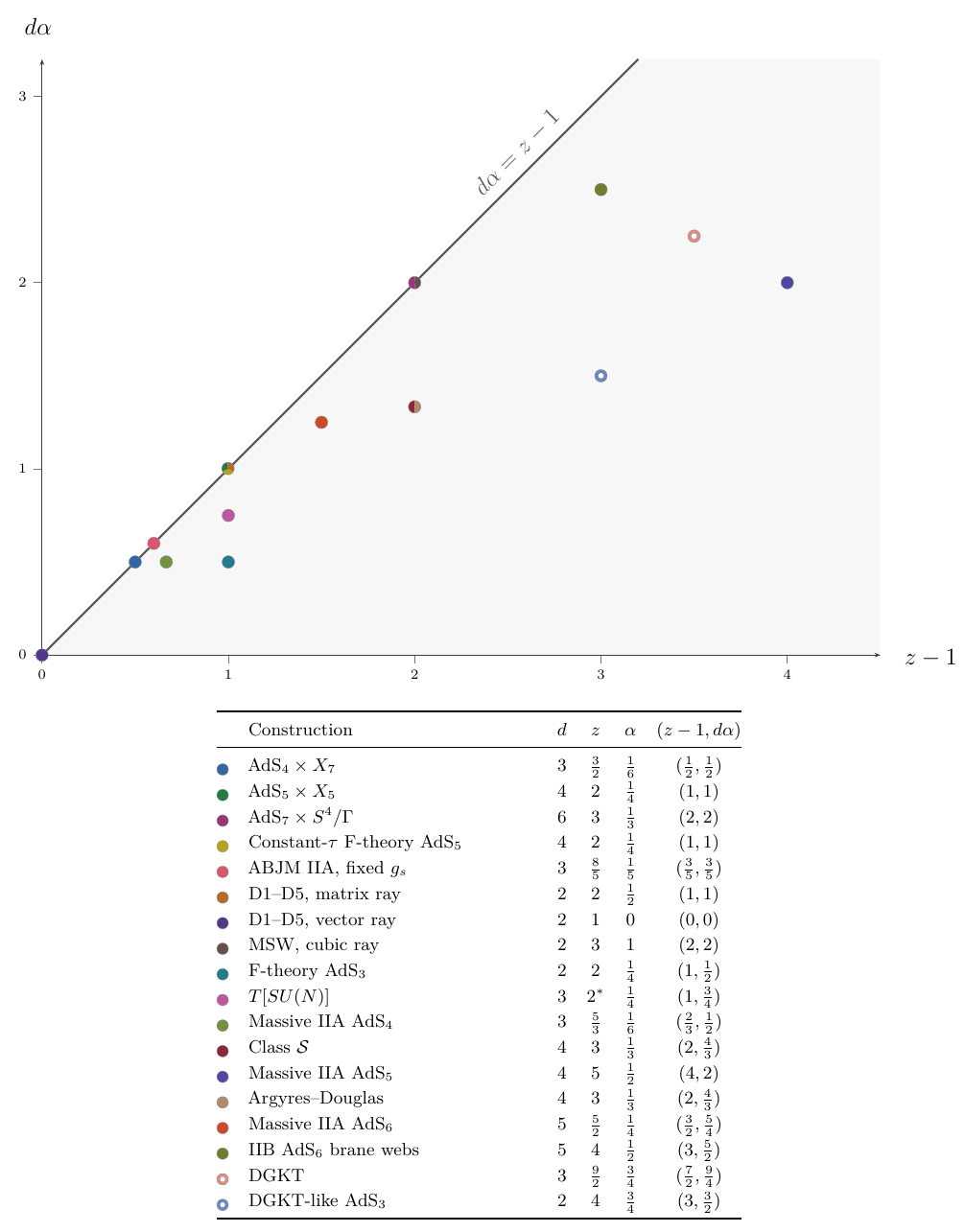} }
\caption{A survey of large $N$ CFT$_d$ central charge and higher-spin gap data, with $c \sim N^z$ and $\Dgap \sim N^\a$. Split-color markers denote exactly coincident coordinates. Several models admit other scaling regimes (all of which lie on or below the diagonal), and we have made simplifying choices here. All known vector models live at the origin. The asterisk denotes a $\log N$ multiplicative factor in the $c$-scaling. The open circles correspond to candidate scale-separated flux vacua, which we address in Section \ref{sec6}. Further explanation and references can be found in the main text and Appendix \ref{appa}.}
\label{survey}
\end{figure}

The observation pertains to $\Dgap$, the {\it single-trace, higher-spin gap} of a large $N$ CFT -- that is, the dimension of the lightest single-trace local primary operator of spin $J>2$. This quantity is known to be an abstract proxy for a ``coupling''. A famous, and conjecturally sufficient \c{HPPS}, hallmark of an emergent Einstein gravity dual is that $\Dgap \rar \i$. We observe that, across known large $N$, unitary CFTs in $d\geq 2$ spacetime dimensions, 
\e{main}{\Dgap^d \lesssim \dc}
where $\dc$ is the central charge increment between successive theories at asymptotically large $N$. The evidence is both holographic and purely field-theoretic. We direct the reader to Figure \ref{survey}, which exhibits many examples of this. A more precise statement, and its family-level extension, will be made below. 

This bound is parametrically tight in many examples, and is stronger than existing bounds on $\Dgap$ \c{CEMZ,Caron-Huot:2022jli}. It is also of a fundamentally different nature: it relates $\Dgap$ to the {\it cross-theory} quantity $\dc$ within a {\it family}. Rewriting the upper bound in terms of $c$, taking power law scaling $c \sim N^z$ for clarity, shows that the growth rate is sensitive to the exponent $z$: letting $\Dgap \sim N^\a$, the bound implies $d\a+1\leq z$. In other words, the precise scaling law of the central charge contains information about the spectrum of the semiclassical theory. This underscores an important lesson of this work: \textit{a collection of apparently individually acceptable quantum gravitational theories may, in the semiclassical limit, have an inadmissible cumulative behavior along a physical sequence.}

In {\bf Section \ref{sec2}}, we present a survey of evidence for \eqr{main}, using a selection of theories from Figure \ref{survey}.

\begin{figure}[t]
    \centering
    \includegraphics[scale=.8]
    {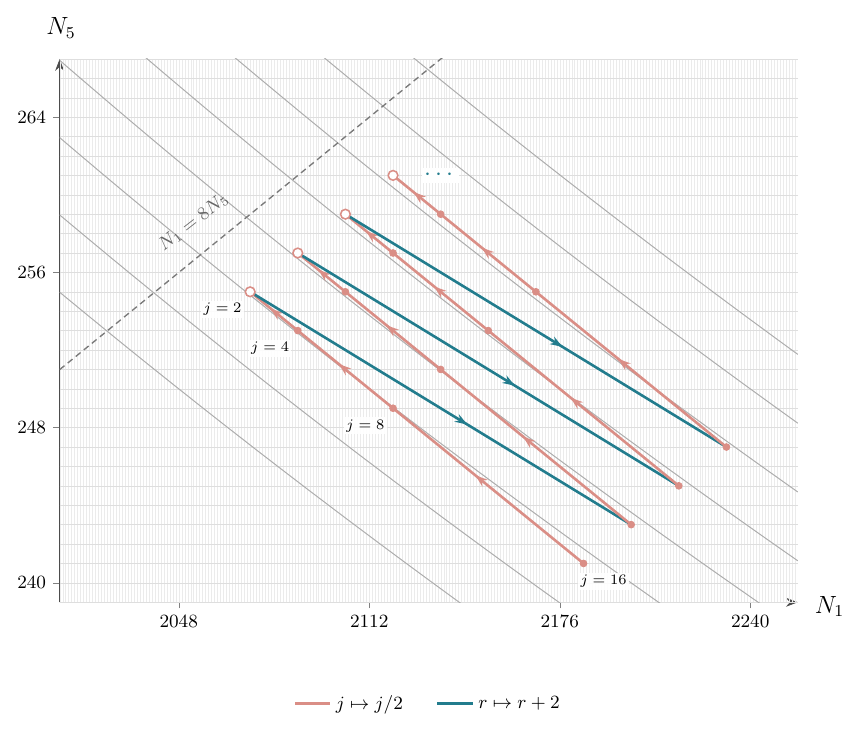}
    \caption{A D1--D5 zigzag with $a=8$, $r=257,259,261,263$,
    and $j=16,8,4,2$ in each block. Coral arrows move within a block;
    teal arrows move to the next block. Central charge increases
    along the path even though individual charges can decrease.}
    \label{zigfig}
\end{figure}

In {\bf Section \ref{sec3}}, we explain the origin of this observation and give a heuristic physical explanation. In short, the relation would be explained by generalized branes, objects that interpolate among CFTs. Their tension sets the scale on the right of \eqr{main}, and upper-bounds the scale on the left. This generalizes the mechanism for \eqr{main} in string/M-theory. We present an apparent counterexample to \eqr{main} using an ``unnatural'' zigzag path through the D1--D5 theory space (see Figure \ref{zigfig}). This suggests a positivity requirement for natural paths, and sets the stage for the concept of families. 

\begin{figure}[t]
\centering
{
\includegraphics[scale=1]{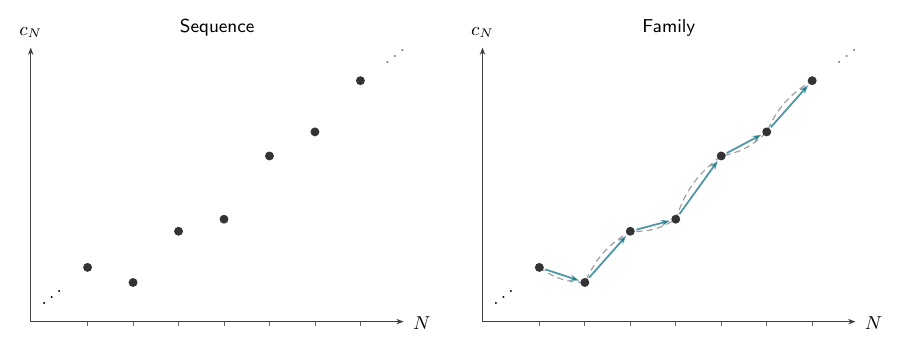} 
}
\caption{A {\it sequence} of CFTs is an arbitrary set labeled by an integer $N$. A {\it family} of CFTs, $\cF$, is a sequence with provenance. The latter is represented heuristically as the dashed line. Changes, shown as arrows, move between successive members of $\cF$ as prescribed by the construction.}
\label{famfig}
\end{figure}

In {\bf Section \ref{sec4}}, we initiate a working definition of families of unitary CFTs in $d\geq 2$. In particular, we present some structural elements that distinguish what one should rightfully call a family, $\cF$, from an arbitrary sequence of theories labeled by an ordinal coordinate $N$. The main distinction is what we call {\it provenance}: a family has an independent construction which precedes enumeration (see Figure \ref{famfig}). Other relational properties flesh out this definition, and we make contact with properties of known CFTs. Let us make very clear that we are not proposing a complete definition of $\cF$, nor any degree of stratification of the space of CFTs into families (if that is even possible). Rather, our treatment is influenced and encouraged by an analogous paradigm in analytic number theory, in which there exists a working definition of {\it families of $L$-functions}, reviewed in Section \ref{s44}: despite its ultimate meaning being incomplete (and an ongoing subject of study), that concept has been very profitable for rigorously probing properties of $L$-functions, and for formalizing what one wants to understand.

\begin{figure}[t]
\centering
{
\includegraphics[scale=1]{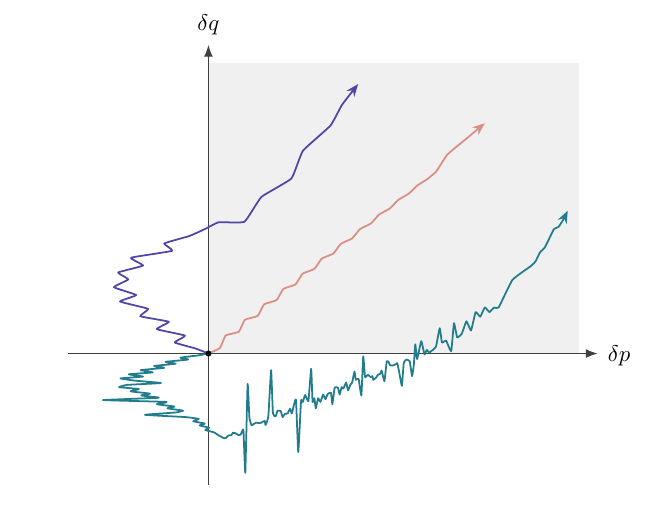}
}
\caption{Ordered families obey eventual positivity, shown here in a plot of charge increments for a family with two charges $(p,q)$. In the pre-asymptotic range of $N$, the behavior is unconstrained.}
\label{qfig}
\end{figure}

\begin{figure}[t]
\centering
{
\includegraphics[scale=1.03]{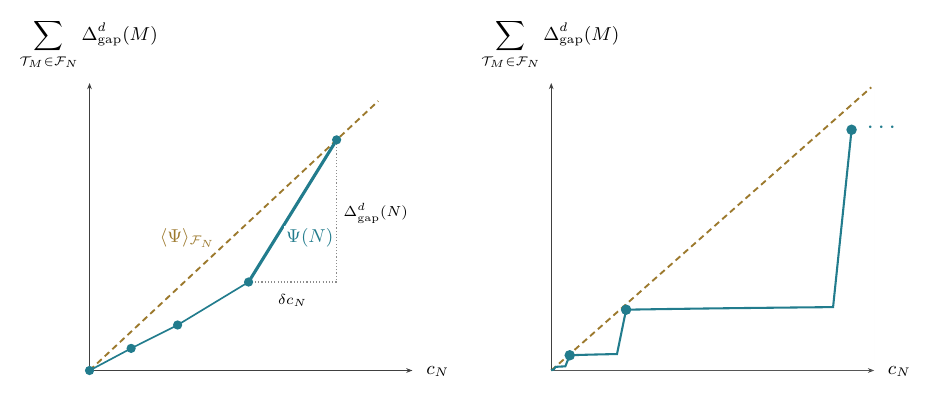} 
}
\caption{{\it Left:} The family gap bound, $\<\Psi\>_\cF<\i$, requires that the slope of the golden dashed line be bounded as its endpoint runs off with $N\rar\i$. The pointwise bound, $\Psi<\i$, requires that the slope of increments be bounded. {\it Right:} An artificial example of a family that would obey $\<\Psi\>_\cF<\i$ but violate $\Psi<\i$, due to erratic spikes at regular intervals.}
\label{v1v5fig}
\end{figure}

In {\bf Section \ref{sec5}}, we formulate the family-level avatar of \eqr{main}. For the large $c$ considerations at hand, we are interested in those families which are $c$-unbounded, and obey an ordering condition on their charge data, eventually in $N$; we call these {\it ordered} families. (In brane language, ordering demands that branes are only added, not removed, at each step at sufficiently large $N$; see Figure \ref{qfig}.) For a theory $\cT_N \in \cF$, introduce the ratio
\e{Psidefintro}{\Psi(N) := {\Delta^d_{{\rm gap}}(N)\over \delta c_N}\,,}
where $\dc_N := c_N - c_{N-1}$ and $\Delta_{{\rm gap}}(N)$ is an effective higher-spin gap asymptotic to $\Dgap$. Whereas \eqr{main} is the step-wise statement
\e{v1intro}{\Psi := \limsup_{N\rightarrow\infty} \Psi(N)<\i\,,}
the family gap bound is (in a slight abuse of notation)
\e{v5intro}{\langle \Psi\rangle_\cF := \limsup_{N\rightarrow\infty} \,\langle \Psi\rangle_{\cF_N}<\i\,,}
where $\<\cdot\>_{\cF_N}$ is defined in \eqr{Fav} as a normalized $\dc_N$-weighted average over a family truncated at ordinal index $N$. This average is, equivalently, the ratio in \eqr{Psidefintro} but with both numerator and denominator summed over the family $\cF_N$. See Figure \ref{v1v5fig}. We view this as a semiclassical constraint on families $\cF$ of quantum theories -- a kind of correspondence principle for AdS gravity. The familiar holographic families, with power law scaling, are {\it self-averaging} in $N$: that is, $\<\Psi\>_\cF = \Psi$. 

In {\bf Section \ref{sec6}}, we grant \eqr{v5intro} and discuss its conditional holographic implications for families of AdS vacua. The aspects covered are clear from the table of contents, but we highlight two here. First, the bound immediately rules out Planckian scale separation for such families. It further provides strong constraints on the degree of allowed scale separation: for instance, for power law families of AdS$_{d+1}$ vacua with $n$ large extra dimensions,\foot{Throughout the paper, $N$ is always an ordinal coordinate of a family or subfamily. Any identification with a rank, flux, brane number or other constructional coordinate is stated separately. We use the following asymptotic notations: $x(N)\sim y(N)$ means agreement in scaling, $x(N)\lesssim y(N)$ means an upper bound in scaling, and $x(N)\approx y(N)$ means agreement in both scaling and leading-order coefficient.}
\e{gpowerintro}{{\frac{M_{\rm HS}}{M_{p,d+1+n}}\lesssim N^{\g(d,n,z)}\,,\quad \g(d,n,z) := \frac{(n-1)(z-1)-d}{d(d-1+n)}}}
We unravel the consequences of this novel bound. Second, in Section \ref{s64} we introduce the concept of {\it extremal families}, defined as those which maximize the higher-spin scale subject to \eqr{gpowerintro} and the Planck scale of gravitational effective field theory in $D=d+1+n$ dimensions. Their exponents $(z,\a)$ obey \eqr{ext}. Moreover, they are constrained by the integrality of spacetime dimension to be {\it rational}. We explain the $z=d/2$ pattern of the canonical AdS/CFT dual pairs in these terms.

Several appendices complement the main text.

\sec{Evidence}\label{sec2}

We begin by presenting evidence for \eqr{v1intro} from both holographic and non-holographic constructions. All of the examples of this section have regular power law scaling of $c$ and $\Dgap$, which we parameterize as
\e{power}{c \sim N^z\,,\quad \dc \sim N^{z-1}\,,\quad \Dgap \sim N^\a}
as $N\rar\i$. We will typically omit $O(1)$ constants, in order to focus on scaling. 

The essential content of this section can be seen in Figure \ref{survey}. We discuss here only a selection of the theories in the table. References for each entry may be found in Appendix \ref{appa}. We will be brief and informal because the theories are familiar, and the calculations, elementary: indeed, what follows essentially amounts to checking whether $d\a+1\leq z$.

\ssec{Canonical strongly coupled large $N$ CFTs}

We first consider the most canonical cases of AdS/CFT dual pairs, beginning with $d=4$, $\cN=4$ super Yang-Mills, dual to AdS$_5 \x S^5$. At large $N$, one has $c\sim N^2$, so $\dc \sim N$. On the other hand, the higher-spin gap is set by the string scale in AdS$_5 \x S^5$, as was independently confirmed from integrability \c{GromovKazakovVieira,GromovQSCStrongCoupling}: in holographic terms, $\Dgap \sim {L/ \ell_s} \sim (g_s N)^{1/4}$, where $g_s := g_{\rm YM}^2$ is the ten-dimensional string coupling. Thus, $\Psi = O(g_s)$, and $d\a + 1 = z$.  

In the $d=6$, $(2,0)$ CFT dual to AdS$_7 \x S^4$, one has $c \sim N^3$. The higher-spin gap is set -- or, in principle, merely upper-bounded (having not yet been established by independent CFT methods) -- by the 11-dimensional Planck scale, $\Dgap \lesssim {L/ \ell_{p,11}} \sim c^{1/9}$. Thus, $\Dgap^6 \lesssim \dc$, and hence $\Psi = O(1)$. Likewise, in the ABJM theory dual to AdS$_4 \x S^7/\Z_k$ at fixed $k$, one has $c \sim k^{1/2}N^{3/2}$ and $\Dgap \lesssim c^{1/9}$. Thus, $\Dgap^3 \lesssim \dc$ and hence $\Psi = O(1)$.

In the AdS$_5$ and AdS$_4$ cases, replacing the spheres by compact Einstein spaces rescales $c$ and $\Dgap$ by $O(1)$ volume ratios, leaving the scaling of $\Psi$ intact.  This holographically establishes $\Psi = O(1)$ for the corresponding infinite families of SCFTs with Sasaki-Einstein compact spaces. For example, for an AdS$_5 \x SE_5$ Freund-Rubin background with fixed $SE_5$,
\begin{equation}
c=\frac{N^2}{4v_5}\,, \quad\frac{L^4}{\alpha'^2}=\frac{4\pi g_sN}{v_5}\,, \quad \text{where}\quad v_5 := {\Vol(SE_5)\o \Vol(S^5)}\,.
\end{equation} 
So $v_5$ cancels entirely in $\Psi$. Notice for these cases that the scaling $\Psi=O(1)$ is much stronger than the CEMZ bound in $d=4$ \c{CEMZ},
\e{cemz}{\qquad \qquad\qquad \Dgap^2 \lesssim \left|{c\o a-c}\right|\qquad \text{\small(CEMZ)}}
with an (explicit \c{Caron-Huot:2022jli}) $O(1)$ constant suppressed: for smooth $SE_5 \neq S^5$, one has $|a-c| = O(1)$, so \eqr{cemz} predicts $\Dgap^2 \lesssim c$, which is much weaker than the actual gap obeying $\Dgap^4 \sim \dc$.  

Indeed, $\Psi = O(1)$ is a property of Freund-Rubin compactifications in general: in AdS$_{d+1} \x X_n$ where $L_{X_n} \sim L_{\rm AdS} \sim L$, flux quantization gives
\e{}{N \sim (LM_{p,D})^{n-1}}
with $D:=d+1+n$, which implies 
\e{}{c \sim (LM_{p,D})^{D-2} \sim N^{1+{d\o n-1}}\,,\quad \Dgap^d \lesssim (LM_{p,D})^d \sim \d c\,,}
as claimed. 

All of the above examples admit discrete orbifolds, which leaves $\Psi$ invariant (a similar cancellation occurs for the deficit parameter of the F-theory D3 backgrounds of \c{AharonyTachikawa}).

An example that is slightly more involved is given by the ABJM theory in the type IIA regime, dual to AdS$_4 \x \mathbb{CP}^3$. In fact, let us treat the entire family of ABJM theories with $SU(N)_k \x SU(N)_{-k}$ gauge group\foot{Generalization to unequal ranks \c{ABJ} or levels \c{GaiottoTomasielloRomans} is straightforward.} that interpolates between the M-theory ($k=1$) and type IIA supergravity ($k \sim N$) regimes. We parameterize the $N$-scaling of $k$ as 
\e{}{k \sim N^\g\,,\quad \g \in [0,1]}
Bounding $\Dgap$ by the respective Planck and string scales in 11 and 10 dimensions gives
\e{}{\Dgap \lesssim N^{\a(\g)}\,,\quad \text{where} \quad \a(\g) = \begin{cases}
\dfrac{1+\gamma}{6},&\quad 0\leq\gamma\leq\dfrac15~~~\text{(M-theory)}\\[6pt]
\dfrac{1-\gamma}{4},&\quad \dfrac15\leq\gamma<1~~~\text{(IIA)}
\end{cases}
}
The crossover from M-theory to type IIA string theory occurs at $\g=1/5$, which corresponds to fixed type IIA string coupling $g_s \sim (N/k^5)^{1/4}$. On the other hand, 
\e{}{c \sim N^{3+\g\o 2}}
Thus one has
\e{}{\Psi \sim \begin{cases} O(1)\,,&\quad 0\leq\gamma\leq\dfrac15 ~~~\text{(M-theory)}\\[6pt]
N^{1-5\g\o4}\sim g_s\,,&\quad \dfrac15\leq\gamma<1~~~\text{(IIA)}
\end{cases}
}
This is consistent with the bound, and gives the first example of a background for which $\Psi$ decays to zero: as indicated, this is precisely due to the vanishing type IIA string coupling as the M-theory circle shrinks. 

Consider now the class $\cS$ theories dual to AdS$_5$ M-theory backgrounds in which M5-branes are wrapped on punctured Riemann surfaces $\Sigma_g$. The M5-brane origins lead, for fixed $g$ and fixed puncture data, to $c\sim N^3$, whereas the higher-spin gap is (ostensibly) upper-bounded by the 11-dimensional Planck scale, $\Dgap \lesssim N^{1/3}$. However, this being a four-dimensional CFT, one has $\Psi \sim N^{-2/3} \rar 0$, as opposed to $\Psi = O(1)$. The underlying reason for this decay, to be elaborated upon in the next section, is that this is a {\it wrapped brane} background, in which $\Sigma_g$ acts as a sink for M5-brane tension. A similar phenomenon occurs for warped backgrounds, orbifold backgrounds, or orientifold backgrounds, as catalogued in Figure \ref{survey}: for all of these, $\Psi \rar 0$ as $N\rar\i$.

Note that the plot contains two points corresponding to candidate scale-separated AdS$_3$ and AdS$_4$ DGKT-type vacua. The fully non-perturbative existence of these vacua is less well-established. Nevertheless, we observe that they obey the bound. We return to these, and other candidate scale-separated vacua, in Section \ref{sec6}. 

\ssec{D1--D5 CFT}
Let us now treat the D1--D5 system, which is a bit richer.\foot{An analogous analysis may be done for the MSW CFT \c{MSW}, which has a three-charge space and admits sub-cubic paths (Figure \ref{survey} only shows the cubic one).} In particular, this allows us to demonstrate the sensitivity of $\Dgap$ to the particular choice of path through theory space. The D1--D5 theory space is parameterized by two independent positive integral charges $(N_1,N_5) \in \Z_+^2$, as well as many continuous moduli (whose number depends on the choice of $M_4$ in the bulk dual). The central charge is
\begin{equation}
c(N_1,N_5)=6N_1N_5+c_0,
\label{eq:D1D5-c}
\end{equation}
where the charge-independent $c_0$ cancels from all increments. For concreteness, we choose $M_4 = T^4$ in the following presentation, so $c_0=0$. The AdS/CFT dictionary in the type IIB supergravity regime dictates that, on the locus with vanishing RR scalars, the D1--D5 frame obeys
\begin{equation}
\frac{L^2}{\alpha'}=g_{10}N_5=g_6\sqrt{N_1N_5},\qquad
v_4 := \frac{\Vol(T^4)}{\alpha'^2}=\frac{N_1}{N_5},\qquad
g_{10}^2=g_6^2\frac{N_1}{N_5}.
\label{eq:D1D5-RR-frame}
\end{equation}
Weak curvature and weak ten-dimensional coupling in this patch require small $g_{10}$ and large $N_5$. The size of the torus in string units determines whether a six-dimensional supergravity description ($N_1 \gtrsim N_5$) or a ten-dimensional supergravity description ($N_1 \gg N_5$) holds. Altogether, weakly-coupled supergravity requires the parametric hierarchy
\e{RRsugra}{g_{10}\ll1 \ll N_5 \lesssim N_1}
See e.g. \c{AU, AharonyFrumkinMehl} for the S-dual description.

Now, the D1--D5 CFT has a high-dimensional conformal manifold $\cM_{\text{D1--D5}}$, and backgrounds with the same product $N_1N_5$ can be related by U-duality. Therefore, to determine the maximal higher-spin gap, one should extremize over U-duality orbits. In particular, as analyzed originally in \c{Larsen:1999uk} and reviewed in \c{MMT, AU, AharonyFrumkinMehl}, for fixed $N_1N_5$, the number and position of strongly coupled cusps of $\cM_{\text{D1--D5}}$ are determined by the coprime factorizations into $(N_1,N_5)$. In the neighborhood of the perturbative RR supergravity point \eqr{RRsugra}, 
\e{d1d5gapbound}{\Dgap^2\lesssim \min(N_1,N_5)}
as inferred from the bulk string scale. 

However, even before any further quantitative analysis using the U-duality group, one can already draw some valuable conclusions, as presaged in the Introduction. By the basic elements of the AdS/CFT dictionary recalled just above, a supergravity point on moduli space -- that is to say, a point with large higher-spin gap -- requires $N_1\gtrsim N_5 \gg 1$. From the CFT point of view, in which we are given a family of these CFTs parameterized by the central charge $c$, if $c/6$ does not admit a factorization into such a pair $(N_1,N_5)$, the theory will simply not have a large-gap limit {\it anywhere} on $\cM_{\text{D1--D5}}$. The case of prime $c/6$ is the extreme scenario, but any value of $c/6$ for which its constituent $(N_1,N_5)$ cannot both become parametrically large must have bounded $\Dgap$.\foot{It is a very interesting challenge from the D1--D5 symmetric orbifold point of view to ``see'' that not any large $c$ path through charge space admits a supergravity point on moduli space. To put it conversely, if a given calculation in the symmetric orbifold is not sensitive to the primality properties of $c/6$, then it cannot claim to detect behavior that identifies a supergravity point putatively connected by an exactly marginal deformation. This presents a certain principled obstacle to attempts at interpreting perturbative orbifold computations holographically.} More precisely, we mean this in the asymptotic sense of a family for which $c/6 \rar\i$ becomes unbounded. 

This leads us to distinguish between two natural paths through the D1--D5 moduli space:
\es{}{\textbf{Matrix ray:}&\quad (N_1,N_5) \approx (N,aN)\,,\quad \,N\rar\i\,,~a~\text{fixed}\\
\textbf{Vector ray:}&\quad (N_1,N_5) \approx (N,q)\,,\qquad N\rar\i\,,~q~\text{fixed}}
The names come from the respective central charge scalings. These relations need only be imposed eventually in $N$.\foot{Subleading corrections can enforce coprimality. For example, the matrix ray may be $(N,aN+1)$ where $a\in\Z_+$.} The matrix ray is the usual holographic limit. The vector ray is not usually discussed, but is equally available. Both rays have $\Psi = O(1)$:
\es{}{\textbf{Matrix ray:}&\quad \dc_N \approx 12aN\,,\quad \Dgap^2 \lesssim N\min(1,a) \quad \Longrightarrow \quad \Psi = O(1)\\
\textbf{Vector ray:}&\quad \dc_N \approx 6q\,,\qquad~ \Dgap^2 \lesssim q ~~\quad\qquad \qquad  \Longrightarrow \quad \Psi = O(1)}
These are upper-bounds on $\Dgap$ (a different path through moduli spaces as one varies $N$ could lower the gap, and hence $\Psi$). Here we have focused on the matrix and vector scalings, but later, in Section \ref{sec3}, we will analyze general paths through $(N_1,N_5)$ charge space for which $c\rar\i$.

The above example nicely demonstrates two points made in the Introduction. First, the central charge exponent $z$ of a chosen ray through theory space contains dynamical information about the large $c$ limit theory. And second, the vector ray has bounded $\Dgap$. 

\ssec{Vector models}

Indeed, consider more broadly the case of vector models. As noted in the Introduction, this term does not have a standard abstract meaning in CFT. Let us {\it define} large $c$ vector models as families of theories $\{\cT_N\}$ for which $c(N)$ scales linearly in $N$ as $N\rar\i$. By this definition, vector models have $\dc_N = O(1)$. If the proposed bound $\Psi<\i$ holds, it implies that they must have $\Dgap = O(1)$ as well. 

Our empirical observation is that this is true for all known vector models. There are essentially two types of vector models for the sake of our classification. 

The first class, which includes the famous Wilson-Fisher and Gross-Neveu(-Yukawa) models in $d=3$, are ``slightly broken higher spin theories'' \c{MZ}. These CFTs have a nearly-exact higher-spin symmetry: $\Dgap$ saturates the unitarity bound up to $1/N$ corrections, due to an infinite tower of almost-conserved currents. Souped-up versions of these theories involve fundamental matter coupled to Chern-Simons gauge groups \c{AharonyGurAriYacobyBosons, GiombiMinwallaPrakashEtAl, AharonyGurAriYacobyBosonization, GurAriYacobyFermions, Aharony:2012ns}, and an ABJ embedding thereof \c{ChangMinwallaSharmaYinABJ}. The ABJ story is especially clean: with a $U(N)_k \x U(M)_{-k}$ gauge group, interpolating between $N\approx M\gg1$ and $N \gg M =O(1)$ regimes exhibits a crossover from $c \sim k^{1/2}N^{3/2}$ and large $\Dgap$ to $c \sim N$ and $\Dgap = O(1)$, with $\Psi = O(1)$ in both cases. In $d=2$, the $W_N$ minimal models admit a large $N$, 't Hooft-like limit in which $c \sim N$ \c{GG}; they also carry infinitely many conserved Virasoro primary higher-spin currents. 

The second class are those vector models in which $\Dgap$ is not close to the unitarity bound. We already discussed the D1--D5 vector models above. One can also consider class $\cS$ theories in the limit of fixed $N$ but large genus $g$: the central charge is linear in $g$, and indeed, this is not a holographic limit. An instructive class of vector model CFTs is that of IR fixed points of disordered UV models -- so-called ``disordered vector models'', or ``SYK-like models.'' These were first written down by \c{Murugan:2017eto} in $d=2$ in an attempt to find higher-dimensional generalizations of SYK, with later versions in both two \c{Chang:2019yug, ChangShenDisordered} and three \c{ChangColinEllerinPengRangamaniStrings,CP3d} dimensions. All of these theories are vector model CFTs. There exist versions both with and without supersymmetry. In the original \c{Murugan:2017eto} model, for example, the IR SCFT has $c={3N\o 2}(1-{2\o \hat q})$, and flows from a theory of $N$ superfields with a $\hat q$-fold interaction term. As for $\Dgap$, these papers do not actually compute $\Dgap$ directly, but rather, they compute the Lyapunov exponent, $\l_L$, for out-of-time-order four-point functions. In $d\geq 2$ CFTs, this is a different abstract proxy for the ``coupling'' of a CFT (its relation to $\Dgap$ is not known in general). The chaos bound of \c{MMS} says that, in a unitary CFT at inverse temperature $\b$, 
\e{}{0 \leq {\b\l_L\o 2\pi} \leq 1}
where the upper bound is saturated by Einstein gravity. In the model of \c{Murugan:2017eto}, ${\b\l_L\o 2\pi}$ takes strictly positive values for bounded $2<\hat q<\i$, and is maximized by 0.5824 at $\hat q=3$. Similar behavior, in which $\l_L$ takes intermediate values, was found in the other models. Thus, if we accept the implication of the HPPS conjecture that
\e{}{\l_L <{2\pi\o\b} \quad \Longrightarrow \quad \Dgap <\i\,,}
then the disordered vector models all have $\Psi<\i$. 

This concludes our brief evidence review. The reader may readily confirm the other examples in Figure \ref{survey} using the references in Appendix \ref{appa} (as well as yet other examples we have not included). 

\sec{Why the bound holds}\label{sec3}

The evidence for $\Dgap^d \lesssim \dc$, culled from both holographic and non-holographic examples, invites a more fundamental explanation. Why should the higher-spin gap know about the change in degrees of freedom between neighboring theories? And why should that change respect a parametric bound?

The survey suggests two broad patterns. In Freund-Rubin examples, $\Psi = O(1)$. In wrapped, warped, and brane-web constructions, $\Psi \rar 0$ with a power of $N$. There is a physical picture behind this difference: wrapping can increase the {\it effective brane tension,} while warping or additional light sectors can lower the higher-spin gap. For example, a D-brane wrapping an internal cycle down to $d$ worldvolume dimensions has
\e{}{T_{\rm brane}\sim\frac{v}{g_s}M_s^d\,,}
where $v\gtrsim 1$ includes any multiplicity of wrapped volume in string units.\foot{Other extended objects may come with different powers of $g_s$.} Internal structure provides, loosely speaking, a {\it tension sink} for a higher-dimensional brane: increasing the ``cost'' of the brane need not raise the higher-spin gap by the corresponding amount. Upper-bounding the higher-spin scale $M_{\rm HS}$ by the string scale, we may write this as
\e{}{\frac{M_{\rm HS}^d}{T_{\rm brane}}\lesssim\frac{g_s}{v}\,.}

What does this have to do with $\dc_N$? The key point is that branes in quantum gravity also play the role of interpolating between AdS vacua, or in CFT language, interpolating from $\cT_{N-1}$ to $\cT_N$. The semiclassical holographic identification is the parametric relation, in AdS units,\foot{Here and below, $\dc$ really means $\dc_N := c_N - c_{N-1}$, and likewise for $\Dgap(N)$, but for the purposes of this discussion we suppress the $N$ indices (to be restored later).}
\e{}{T_{{\rm brane}}\sim\dc}
In the canonical conformal brane backgrounds, this is the tension of a single elementary brane, whereas in more complicated backgrounds generated by multi-brane configurations, the tension $T_{{\rm brane}}$ is to be viewed as that of an effective brane. 

The Euclidean path integral gives this relation a further interpretation. In the near-horizon description, the corresponding objects become domain walls across which flux changes, separating bulk regions associated with the two theories $\cT_{N-1}$ and $\cT_N$. A wall with worldvolume $\Sigma$ contributes a tension term to the action
\e{}{S_{\rm brane} \supset T_{\rm brane}\int_\Sigma d^dx\,\sqrt{\gamma}\sim T_{\rm brane} }
when its extent is of order the AdS radius. The semiclassical contribution to the path integral is proportional to $e^{-S}$. Assuming charge couplings do not cancel the brane term, we see that the tension $T_{\rm brane}$, and hence the central charge increment $\dc_N$, naturally measures the cost of a unit step in theory space. 

Altogether, this suggests the following generalized brane interpretation of $\Psi$,
\e{}{\Psi=\frac{\Dgap^d}{\dc}
    \sim\frac{M_{{\rm HS}}^d}{T_{{\rm brane}}}}
In physical terms, the content of the bound $\Psi<\i$ is the statement that {\it the higher-spin scale is set by the scale of effective brane excitations.} The exponent $d$ is simply dimensional analysis. The internal contributions $v$ to effective tension, discussed above, explain when and why many constructions give $\Psi\rar0$.\foot{The numerator also admits an interpretation in terms of bulk resolution. An AdS$_{d+1}$-sized spatial region contains of order $M_{\rm HS}^d$ independent one-particle modes of a light propagating field below the higher-spin scale. For $d>2$, the graviton itself supplies such a count. Then $\Psi<\i$ says that the tension of the brane per unit volume cannot vanish at the EFT scale, else it would invalidate the EFT. A somewhat similar logic was given in the top-down analysis of \c{CPVR} which, motivated by BPS flux compactifications, used AdS domain walls to propose a bulk bound $\Lambda_{\rm UV}^d\lesssim (M_{p,d+1})^{d-1}$, for some choice of UV cutoff. If we further identify $\Lambda_{\rm UV}$ with $M_{\rm HS}$, translating holographically gives $\Dgap^d\lesssim c$. This bounds by $c$ instead of $\dc$ (and is hence strictly weaker for any power law family).} We may describe this picture as a kind of {\it brane universality}, with a specific spectral imprint. 

That the asymptotic $\Dgap$ should be viewed as a property of a {\it family} of theories is a central lesson of this discussion. Accordingly, it provides an informal hint of distance in theory space: along a regular power-law ray \eqr{power}, the exponent $z$ records how ``close'' successive theories are to one another in central charge. Note, in particular, that vector models, with $z=1$, are the edge case: the theories are ``too close'' to one another to develop a large gap at large $N$, because the effective branes are tensionless.

Motivated by this picture, the remainder of the paper is devoted to taking seriously these notions of generalized branes and families of CFTs, such that we may understand the higher-spin gap bound from a first principles CFT point of view. 

\ssec{An apparent counterexample: D1--D5 zigzags and positive paths}\label{s31}
The obvious first question, then, is whether the set of known examples treated above share a special property that might be violated by a suitably ``strange'' path through theory space. Indeed, we now present an example from the D1--D5 CFT, depicted in Figure \ref{zigfig}, in which $\Psi \rar\i$ as $c\rar\i$.

Fix a positive power of two, $a:=2^s$, and consider
   \e{eq:d1d5-zigzag}{ N_1=a(r+j),\qquad N_5=r-j,\qquad
    c(r,j)=6a(r^2-j^2)\,,}
where $r$ is sufficiently large and odd, and $j=2^t \leq \sqrt{r}$ with $t\in\Z_+$. The prescription is as follows: within each $r$-block, visit the allowed values of $j$ in decreasing order, then move to the next odd $r$. It is straightforward to check that the central charge strictly increases with each step, and every charge pair is coprime. Moreover, $N_1/N_5=a(1+O(r^{-1/2}))$, so the path remains close to a matrix ray.

For the step $j=2^t\to2^{t-1}$,
\begin{equation}
    (\delta N_1,\delta N_5)=(-2^{t-1}a,2^{t-1}),
    \qquad
    \delta c=18a\x 4^{t-1}.
\end{equation}
For every $r$, the set of allowed values of $t$ includes order one values, independent of $r$ (e.g. we may consider the $t=2$ step). For such steps, the central charge increment is of order one, for every $r$. At $r\rar\i$, the central charge diverges as $c\sim 6ar^2 \rar\i$. In a region of the moduli space where $\Dgap$ attains the maximum, we have $\Dgap^2 \sim r$. Putting things together, we see that along this zigzag path, there is an infinite subsequence on which $\Psi\sim r \rar\i$. 

The mechanism is simple: adding one kind of charge while {\it removing} another allows cancellations in $\delta c$ that preserve its positivity, without a corresponding reduction in the gap. In D-brane language, the above moves are not merely adding D5-branes, but also taking D1-branes away (the coral moves in Figure \ref{zigfig}); every now and then, the two swap (the teal moves in Figure \ref{zigfig}).

On the other hand, for {\it charge-ordered} paths -- that is, paths where every move is, at least eventually in $N$, adding both D1 and D5 charge -- those cancellations are absent. It is elementary to show that $\Psi$ remains bounded along ordered paths. Assume the estimate \eqr{d1d5gapbound} holds uniformly in the region under consideration. For any distinct successive integral charge pairs with $N_1\ge p_-$ and $N_5\ge q_-$,
\begin{equation}
    N_1N_5-p_-q_-\ge\min(N_1,N_5), \qquad \delta c\ge6\min(N_1,N_5), \qquad \Psi = O(1).
\end{equation}
Indeed, increasing $N_1$ costs at least $N_5$ in the product $N_1N_5$, and vice-versa. This proof says nothing about the nature of the paths, which may be arbitrarily jagged, anisotropic, or thin. 

This suggests an important ordering requirement for $\Psi<\i$ more generally, supported by the physical brane point of view: at sufficiently large $N$, we should only add branes. We would like to formulate this intrinsically in CFT language. This segues to the notion of families of CFTs. 

\sec{Families of CFTs}\label{sec4}

We will be concerned with collections of CFTs $\{\cT_N\}$, where $N\in\Z_+$ is an ordinal coordinate labeling a given CFT $\cT$. We work in the space of unitary, local CFTs in $d\geq 2$ dimensions. We fix a dimension $d$, and a subspace with fixed reference conformal or superconformal algebra. For $d=2$, this is to be understood as fixing reference left- and right-moving local conformal symmetry algebras, $\cA \x \overline{\cA} \supseteq \text{Virasoro} \x \overline{\text{Virasoro}}$. By ``reference'' we mean that some theories in the subspace may acquire an enhanced symmetry; in $d=2$, for example, the maximal chiral algebra may vary. Generalized global symmetries may also vary. We assume for simplicity that each theory has a unique stress tensor, though this could plausibly be relaxed. 

The stress tensor two-point function is canonically normalized by a constant, $C_T$. Since degrees of freedom are counted by the $a$-type conformal anomaly coefficient in even dimensions and the sphere free energy in odd dimensions, the ``$c$'' to which we refer throughout the paper may be best thought of as identified with those quantities. In practice, however, this is immaterial to our large $N$ analysis -- and indeed, bounds in even dimensions \c{HM} constrain these quantities to have the same large $N$ scaling. For further discussion of these aspects, see Appendix \ref{appg}.

\ssec{Provenance}

At its core, a family $\cF$ is a sequence $\{\cT_N\}$ that is supplied by a construction. In contrast to an arbitrary sequence of CFTs, we say that a family has {\it provenance}, which we define as follows.

\begin{quote}
\noindent\hspace*{-0.75em}%
\begin{tabularx}{\dimexpr\linewidth+1.5em\relax}
  {@{}l@{\hspace{0.4em}}X@{}}
\textit{Provenance}: &
membership and parametrization are determined by constructional data defined independently of the enumeration of the members.
\end{tabularx}
\end{quote}

\noindent Altogether, a CFT formation supplies the triple
\begin{equation}
(\cF,\sP,Q),
\qquad
Q:\cF\longrightarrow\sP,
\label{eq:formation}
\end{equation}
where $\cF$ is a family of pairwise inequivalent CFTs, $\sP$ is the full set of data carried by that construction, and $Q(\cT)$ records those data for $\cT\in\cF$. In a gauge-theory family, $\sP$ may contain ranks, gauge groups, matter representations, couplings, discrete data, and so on. In a holographic brane construction, it may contain charges, fluxes, transverse geometry, moduli, and so on. These data enable provenance. See Figure \ref{famfig}. 

A constructional parameter of $\sP$ has a ``life of its own'': the construction of $\cF$, the meaning of $\sP$, and the map $Q$ are stated independently of any enumeration of $\cF$. The distinction is not merely between an indexed and an unindexed set. Given any injective list, one can always declare $Q_{\mathrm{ord}}(\cT_N):=N$. This is a parameterization, but it contains no constructional information. Rather, the distinction is between an index {\it induced} by data that defined the family, and an index assigned solely to enumerate preselected members. 

Once $(\cF,\sP,Q)$ has been supplied, let
\begin{equation}
\gamma(\cF):\N\longrightarrow\cF
\label{eq:gamma}
\end{equation}
be an injective sequence. This carves out a {\it subfamily}: $\im\gamma(\cF)$ is a locus of CFTs belonging to $\cF$. It will be useful to introduce the subfamily notation
\e{Fsub}{\cF_\g := \im\g(\cF)}
Its trajectory in constructional data is
\begin{equation}
Q\circ\gamma:\N\longrightarrow\sP
\label{eq:data-path}
\end{equation}
such that $\im(Q\circ\gamma)$ is the corresponding locus of data. Some of the data of $\sP$ may be held fixed in a particular subfamily, while others may vary. 

What we are developing here and below is a working definition. It is not a claimed first-principles classification of all constructional data $\sP$, or all CFTs (reasonably regarded as among the most profound problems in quantum field theory). Note that provenance alone does not require one canonical partition of all CFTs into disjoint families, or finitely many families within a fixed symmetry sector. Different formations may overlap, and the same theory may occur in several independently meaningful families because different constructional data are being held fixed or allowed to vary.

\ssec{Changes}

To flesh out the above definition, guided by physical examples, we would like to identify further physical structure. Let us add to $(\cF,\sP,Q)$ a collection of what one might call constructional changes, $\cW$: for two theories $\cT,\cT' \in \cF$,
\e{changes}{W:\cT \rsa \cT' \in \cW\,.}
By ``constructional changes,'' we mean a relational structure that allows moves between theories, that is prescribed by the provenance. The constructional and relational data have distinct roles: schematically, 
\begin{equation}
\begin{aligned}
(\sP,Q):&\quad\text{locate theories in the family}\\
\cW:&\quad\text{records motion through the family}
\end{aligned}
\end{equation}
The notion of changes $\cW$ is a formal codification of the operation of, say, taking $N \rar N+1$ in an $SU(N)$ gauge theory. It accounts for the fact that connectedness between two theories $\cT,\cT'\in\cF$ is stronger than provenance: being part of the same family is a statement about how theories are defined, not necessarily about how they can be connected to one another by a prescribed set of operations. This is what $\cW$ records.\foot{It is surely possible to make this more mathematically formal, to reduce the ambiguity in what one gets to specify. See e.g. \c{Zeev:2022cnv} for a version of this.} Again, this is more than the bare existence of some abstract arrow system assigned to pairs in a list.

Among these changes $\cW$, we will soon want a notion of positivity, i.e. of ``moving upwards'' along the family (recalling the D1--D5 zigzag example of Section \ref{s31}). We can distinguish a collection $\cW_+$ of {\it positive changes}, closed under composition:
\begin{equation}
\Wp=\{W:\cT\rightsquigarrow \cT'\text{ a positive allowed constructional change}\},
\label{eq:add-Wplus}
\end{equation}
obeying
\e{eq:add-composition}{\forall~W_i:\cT_{i-1}\rightsquigarrow \cT_i \in \Wp\,,~~ W_{i+1}\circ W_i:\cT_{i-1}\rightsquigarrow \cT_{i+1}\ \in\Wp}
Composition has its ordinary associative meaning. We can make things less formal by specializing to the cases we are most familiar with from explicit CFT constructions, in which these changes come with an additive charge structure, which we do now. 

\sssec{Graded provenance: Additive changes}

Relative to the above general setup, canonical examples of families of CFTs have extra {\it graded} structure. It is useful to specialize to cases where the constructional data contain $r$ independent charges, with $r$ fixed along the family. After choosing an origin, write the charge vector as
\begin{equation}
q(\cT)=(q^1(\cT),\ldots,q^r(\cT))\in V
\label{eq:add-lattice}
\end{equation}
The map 
\e{}{Q_V:\mathcal F\longrightarrow V, \qquad V\simeq\mathbb R^r}
is the charge part of $Q$: that is, $\sP$ may contain a charge component, as well as other transverse components with other non-additive data (i.e. $\sP \supseteq V \x \sP_\perp$, where the $\x$ stands for a more general product that allows correlations between the components). 

The additive structure makes it easy to talk about the allowed positive changes $W\in\mathcal{W}_+$: they are simply those with {\it positive charge increments}. Let us define a degree, a vector-valued grading of the changes:
\e{}{\deg W:=q(\mathcal T')-q(\mathcal T)\,,\quad W:\mathcal T\rightsquigarrow\mathcal T'}
A composition rules follows immediately,
\begin{equation}
\deg(W_{i+1}\circ W_i)=\deg W_{i+1}+\deg W_i,
\label{eq:add-degree}
\end{equation}
which implies that for any finite composition $P=W_k\circ\cdots\circ W_1$, deg $P$ is the charge difference of the first and last theories in the chain; thus, degrees are additive and path-independent. Moreover, in this additive setting, positivity is just the condition that the degree vector is semi-positive:
\e{Wposdeg}{\forall \,W\in \cW_+,~~ \deg W\in V_+\setminus\{0\}\,,~~V_+  \simeq \R_{\ge0}^r\,.}
Equivalently, 
\e{Wpdegup}{\forall \,W\in \cW_+,~~ \text{deg $W \succeq 0$,~~ deg $W \neq 0$}\,.}
%
In other words, $W \in \cW_+$ if and only if no charges decrease and at least one charge increases.

We note that this structure is retained under -- and indeed, keeps track of -- thinning to a subfamily. Letting $n_1<n_2<\cdots$, the change from $\cT_{n_j}$ to $\cT_{n_{j+1}}$ is the composite of {\it all} omitted positive steps, and
\e{eq:add-thinning-degree}{\deg W_{n_j\to n_{j+1}}
=q(\cT_{n_{j+1}})-q(\cT_{n_j})
=\sum_{n=n_j+1}^{n_{j+1}}\deg W_n\in V_+\setminus\{0\}.}
This is the precise sense in which the relational data keep track of the ambient family: the enumeration index of a thinning does not become a new physical unit of charge.\foot{For instance, if a canonical rank-one charge family $\cF$ with $q(\cT) \in \Z_+$ is thinned to squares, 
\e{eq:squares}{q'(\cT)\in\{1,4,9,\ldots\},}
then for selected terms $\cT_{n_k}$, one has $q'(\cT_{n_k})=q(\cT_{n_k})$, as opposed to the ordinal label $q_{\mathrm{ord}}(\cT_{n_k})=k$.} 

\sssec{Charge lattices}
We can further specialize to charge lattices: in the case that physical charges are quantized, $V$ is a lattice $\Lam$. Once a common charge lattice is supplied along the family, choosing an origin $q_0$ and a lattice basis lets us write
\e{}{q(\mathcal T)\in q_0+\Lambda\,,\quad \Lam\simeq\mathbb Z^r}
The constructional data now split as
\e{}{\sP=(q_0+\Lam)\times\sP_\perp}
Now positivity is again \eqr{Wpdegup}, but with integral degrees (charge increments):
\e{}{\forall\, W\in\cW_+,~~\deg W\in \Lam_+ \setminus\{0\}\,,\quad \Lam_+ \simeq \Z_{\geq 0}^r}
This case is familiar from holographic constructions with D-brane charges. So, we have a kind of hierarchy of specializations of family structure: $\cF \supset \cF_{\rm graded} \supseteq \cF_{\rm lattice}$. 

\ssec{Large $c$ families}

The above treatment was intended to be general. To prepare for our large $N$ excursion, we now identify those families that are relevant. 

First of all, as an obvious prerequisite for relating to semiclassical gravity, we will only be interested in those families $\cF$ (and subfamilies $\cF_\g$ thereof) for which both $N\in\Z_+$ and the central charge $c_N$ are unbounded from above:
\e{}{c_{N\rar\i} \rar\i\,.}
We call such families {\it unbounded}. 

We wish to study what we call {\it ordered} families. An ordered family is defined to be a family for which the changes $W$ and central charge increments $\dc$ are {\it eventually positive}:
\e{}{\exists\, N_0~:~\forall N\geq N_0,~ W_N \in \cW_+,~\dc_N>0\,.}
The threshold $N_0$ can, of course, depend on the family. The central charge ordering is natural, and easy to understand. When the data $\sP$ admit additive grading, the condition on changes (recalling the definition of $\cW_+$ in \eqr{Wpdegup}) is that the degree of changes is eventually a semi-positive vector, as in \eqr{Wpdegup}: in the language of charge increments, an ordered family obeys
\e{orderdef}{\d q_N \succeq 0\,,~\d q_N \neq 0~~\text{and}~~\dc_N>0}
for all $N \geq N_0$. We nickname these orderings $q$-ordering and $c$-ordering, respectively. We emphasize the eventual-ness of our definition: at finite $N$, we require no particular behavior. See Figure \ref{qfig}. 

In general, $q$-ordering and $c$-ordering are distinct: neither implies the other. We already saw in the D1--D5 zigzag example of Section \ref{sec3} that $\d c_N>0 \nRightarrow \d q_N>0$. In the other direction, one can find cases where $\d q_N>0 \nRightarrow \d c_N>0$ in the presence of multiple charges. In Appendix \ref{appd}, we give a simple explicit example from $d=4$ superconformal QCD.\footnote{In $d=4$ SQCD, $\Dgap$ stays bounded. In principle, it could be that in all families for which $\Dgap$ can grow large (i.e. holographic families), the dependence of $c$ on charges {\it is} such that $\d q_N>0 \Rightarrow \d c_N>0$ eventually. However, absent such a result, we separate the two conditions explicitly in our definition.} 

In the above definition, $c$-ordering is strict: the increments do not decay to zero at large $N$,
\e{}{\liminf_{N\rar\i}\dc_{N}>0\,.}
This has the useful implication that
\e{}{c_N-c_{N_0} = \sum_{M={N_0+1}}^{N} \dc_M \gtrsim N}
as $N\rar\i$. This is a density condition: for a $c$-ordered family, the number of theories below central charge $C$ is at most linear in $C$. Letting
\e{eq:add-chain-count}{\mathcal N_\gamma(C):=\#\{N:c_N\leq C\}}
for a given chain $\gamma$, an ordered chain obeys
\e{eq:add-liminf}{\mathcal N_\gamma(C)=O(C) \quad\Longleftrightarrow\quad \liminf_{N\to\infty}\frac{c_N}{N}>0}
with $\g$-dependent constants implied. We remind the reader that, here and elsewhere, $N$ is the ordinal coordinate. Note that \eqr{eq:add-liminf} implies unboundedness. For a family with regular, power-law scaling $c_N \sim N^z$ at large $N$, this amounts to 
\e{}{z\geq 1}
If one were to only impose unboundedness, as opposed to $c$-ordering, then $z\in(0,1)$ would be possible. We note in passing that there is no known construction of CFTs for which $z\in(0,1)$, whether we restrict to ordered families or not.\foot{On the other hand, many examples with $z<0$ (not relevant here) come to mind, such as the Virasoro minimal models.} 

It is useful to understand the effect of ordering in a simple ad hoc example. Suppose a two-charge family $\{\cT_{p,q}\}$ with integral grading $p,q\in\Z_+$ were to have central charge
\e{eq:two-param-count}{c(\cT_{p,q})=p+\sqrt2\,q\,.}
Its full charge region contains quadratically many distinct levels below $C$. But under eventual positivity of $\d p$ and $\d q$, every step increases $c$ by at least one; without negative moves, $\dc$ cannot be arbitrarily small. Thus, linear counting is respected. 

\ssec{Guidance from number theory}\label{s44}

The preceding definition of CFT families (and its envisaged utility) receives support from the concept of {\it families of $L$-functions} in number theory.

The crucial point relevant for present purposes, on which we elaborate below, is twofold. First, the definition of a family of $L$-functions is very much incomplete: it remains a working definition \c{Sarnak, SST, SarnakGRH2018}. Second, despite this, thinking in terms of families has led to deeper, rigorous understanding of $L$-functions: their zero statistics, moments, connections to random matrix theory, and various taxonomic phenomena.\foot{For a sample of important references on the topic, see \c{KS, IwaniecSarnakNonvanishing, IwaniecSarnakDirichlet, IwaniecLuoSarnak, IwaniecSarnakPerspectives, Sarnak, SST}.}

A number-theoretic construction supplies a set of objects to which $L$-functions are attached,
\e{NTform}{\sP_{\mathrm{NT}} \ni p
\longmapsto
L(s,\pi_p),}
where $\sP_{\mathrm{NT}}$ is the formation data. The set $\{L(s,\pi_p)\}$ comprises a family, which comes with a specified ordering or exhaustion, often (though not always \c{SST}) given by the analytic conductor \c{IK}, call it $\mathfrak C(\pi_p)$:\foot{The analytic conductor is regarded as a measure of the ``complexity'' of an $L$-function. It depends on the parameters appearing in the functional equation in a simple way.} 
\begin{equation}
\cF_{\mathrm{NT}}(N):= \left\{L(s,\pi_p):p\in\sP_{\mathrm{NT}},
\ \mathfrak C(\pi_p)\leq N\right\}\,.
\label{eq:NT-exhaustion}
\end{equation}
In order for this to be a bona fide family, the data $\sP_{\mathrm{NT}}$ should provide an underlying construction. Often, this can be defined independently by means of some arithmetic, geometric, automorphic, or harmonic objects $\pi_p$ to which the $L$-functions are attached. One may define the asymptotic cardinalities of the families,
\begin{equation}
\cN_{\mathrm{NT}}(N):=\#\cF_{\mathrm{NT}}(N)
\label{eq:NT-count}
\end{equation}
which is used to normalize averages over families of $L$-functions. 

A basic example is supplied by primitive Dirichlet characters. Define
\begin{equation}
\sP_{\mathrm{Dir}}
:=
\{(q,\chi):q\geq1,\ \chi\text{ is a primitive Dirichlet character modulo }q\}.
\label{eq:Dirichlet-parameter-space}
\end{equation}
The attached $L$-functions are
\e{}{(q,\chi)\longmapsto L(s,\chi):=\sum_{n\geq1}\frac{\chi(n)}{n^s}}
The $L(s,\chi)$ are standard $L$-functions of Selberg class, of degree-one, with conductor $q$. The family of all primitive Dirichlet $L$-functions thus admits a natural {\it ordering by $q$}:\foot{The restriction to primitive Dirichlet $L$-functions is useful for injectivity, and hence a CFT analogy, because a character modulo $q$ induced from a primitive character of conductor $q_0\mid q$ produces the same primitive $L$-function data. Passing to primitive characters chooses the intrinsic conductor and prevents the same object from being replicated at every imprimitive modulus.}
\e{}{\cF_{\mathrm{Dir}}(N):=\{L(s,\chi):(q,\chi)\in\sP_{\mathrm{Dir}},\ q\leq N\}.}
The cardinality of the set is known to be given by 
\begin{equation}
\cN_{\mathrm{Dir}}(N) = \sum_{q\leq N}\varphi^*(q)
\label{eq:primitive-character-asymptotic}
\end{equation}
where
\begin{equation}
\varphi^*(q)
:=
\#\{\chi\bmod q:\chi\text{ primitive}\}
=
\sum_{m\mid q}\mu(m)\varphi(q/m).
\label{eq:primitive-character-count}
\end{equation}
Using the Dirichlet series
\begin{equation}
\sum_{q\geq1}\frac{\varphi^*(q)}{q^s}
=
\frac{\zeta(s-1)}{\zeta(s)^2},
\label{eq:primitive-character-series}
\end{equation}
the pole at $s=2$ implies $\cN_{\mathrm{Dir}}(N) \sim N^2$. 

There is, however, no universally applicable {\it intrinsic} definition of families $\cF_{\rm NT}$. In practice, a proposed definition remains provisional, applied to a concrete candidate family ``as it arises'' \c{Sarnak} from an underlying formation. Precise definitions do exist, but at a loss of generality (e.g. \c{SST, KowalskiFamiliesCusp, FouvryKowalskiMichelToroidal}). The structure, or provenance, defining a given family must be specified; this ideally refers to an underlying set of objects to which the $L$-functions are attached (e.g. automorphic forms), but there are weaker versions which also refer directly to a property of the $L$-series themselves (as in e.g. \c{CFKRS,KowalskiFamiliesCusp}).  There is certainly nothing like a canonical partition of all $L$-functions into families. Developing general frameworks for families and understanding their statistical behavior remains an active research direction.

Nevertheless, the notion of families of $L$-functions has already been very useful. What number theorists are often interested in are {\it family statistics}, that is to say, averages of the form
\begin{equation}
\mathbb E_N[F]
:=
\frac{1}{\cN_{\mathrm{NT}}(N)}
\sum_{p:\,\mathfrak C(\pi_p)\leq N}F(\pi_p),
\label{eq:NT-average}
\end{equation}
or a normalized weighted version thereof, where $F(\pi_p)$ is some function of the data of $L(s,\pi_p)$. The shape of the statements in the number theory literature is, schematically speaking, of convergence in law to some limiting behavior as $N\rar\i$,
\e{}{\mathbb E_N[F]~~ \longrightarrow ~~\textsf{(universal limiting law)}}
This includes moments, central values, low-lying zeros, equidistribution theorems, and symmetry types. A central topic of study is the distribution of zeros near the central point $s=\half$ on the critical line: instead of going asymptotically high on the critical line $s=\half+it$ for a {\it single} $L$-function, whereupon one encounters a mean zero density $\sim \log t$, one can instead generate a large zero density at a {\it fixed} location on the critical line by summing over a family with diverging analytic conductor, whereupon the mean zero density diverges as $\sim \log\mathfrak{C}$.\foot{In other words, one may take an asymptotic limit in the $q$-aspect of $\mathfrak{C}$ instead of the $t$-aspect.} 

One of the original conjectures in this regard is the universality conjecture of Katz and Sarnak, which we can summarize as follows: for a given family $\cF_{\rm NT}(N)$, the unfolded zero statistics in the neighborhood of $s=\half$, sampled over the family, converge in law to the eigenvalue statistics of random matrix theory as $N\rar\i$, with symmetry type fixed by underlying structure of $\cF_{\rm NT}(N)$. For a very brief elaboration of this conjecture, see Appendix \ref{appe}. 

\sssec{CFT analogy}
Returning to CFT, the number theory story provides a set of guiding principles for a definition of CFT families $\cF$. An important role in the definitions of both $\cF$ and the number theory families $\cF_{\rm NT}$ is played by provenance: formation precedes enumeration. The CFT counterpart to \eqr{NTform} is 
\e{}{\sP_{\mathrm{CFT}}\ni p \longmapsto \cT(p)}
to which is attached a charge $Q(\cT(p)) = p$. The $c$- and $q$-ordering we introduced for CFT families \eqr{orderdef} mirrors the ordering of $L$-function families using $\mathfrak{C}$.\foot{$c$ is even vaguely similar to $\mathfrak{C}$, as both are measures of ``complexity'' of their respective objects.} This is meant as a guiding analogy, not a literal map. One difference is that, in our CFT families $\cF$, the changes $\cW$ are a stronger structure than what number theorists sometimes require, and perhaps they are not fully necessary. But for present purposes, the important point is not the details, nor the precise content of the Katz-Sarnak conjecture, but rather, the utility of {\it a)} families as organizing structures, even in the context of an incomplete working definition; and {\it b)} family averages and weighted versions thereof.

\sec{A family gap bound}\label{sec5}

We are now ready to state the family-level version of \eqr{main}, which we presented in \eqr{v5intro}. The quantity of interest is $\Psi(N)$, whose definition was given in \eqr{Psidefintro} and is repeated here:
\e{Psidef2}{\Psi(N) := {\Delta^d_{{\rm gap}}(N)\over \delta c_N}\,,}
where we recall our notation for central charge increments,
\e{}{\delta c_N := c_N - c_{N-1}\,, \quad  c_0 := 0}
The boundary condition implies $\dc_1 = c_1$, and in turn, the telescoping condition
\e{dcc}{\sum_{M=1}^N \d c_M = c_N}
In Appendix \ref{appf}, we explain how to define the higher-spin gap in the context of an approach to large $N$. 

Let $\cF_N \subset \cF$ denote the subset of theories in $\cF$ up to ordinal index $N$:
\e{}{\cF_N := \lbrace \mathcal{T}_M \in \cF :  1\leq M \leq N\rbrace\,,}
where $\cF := \cF_\i$. Let us introduce a {\it weighted family average} as follows:
\e{Fav}{\langle f\rangle_{\cF_N} := \sum_{{\mathcal{T}_M \in \cF_N}} w_{M|N} f(M)\,,\quad\text{where}\quad w_{M|N} := {\delta c_M\over c_N}}
The weights $w_{M|N}$ are central charge increments normalized by the total central charge along the path through $\cF_N$, and as such obey
\e{}{\sum_{{\mathcal{T}_M \in \cF_N}} w_{M|N} = 1}
In view of \eqr{dcc}, one may also view $\langle f\rangle_{\cF_N}$ as a normalized average of $f(M)$ with weight $\dc_M$. Recall that this weight is positive for families of interest.

We now consider the family average of $\Psi(N)$. Expanding the formula,
\e{}{\langle \Psi\rangle_{\cF_N} = \sum_{\mathcal{T}_M \in \cF_N} {\Delta_{{\rm gap}}^d(M)\over c_N} = {\sum\limits_{\mathcal{T}_M \in \cF_N}\Delta_{{\rm gap}}^d(M)\over  \sum\limits_{\mathcal{T}_M \in \cF_N} \delta c_M}} 
This averaged generalization of $\Psi(N)$, in which both the numerator and denominator are summed over the family, is precisely what we want to consider. The step-wise gap bound on $\Psi(N)$ is
\e{v1}{\Psi := \limsup_{N\rightarrow\infty} \Psi(N)<\i}
The family-level avatar of \eqr{v1} is that, at large $N$, the family average of $\Psi(N)$ is also bounded:
\e{v5}{\langle \Psi\rangle_\cF := \limsup_{N\rightarrow\infty} \,\langle \Psi\rangle_{\cF_N}<\i\,.}
This extends to admissible subfamilies $\cF_\g \subseteq \cF$, as defined in \eqr{Fsub}. 

The condition \eqr{v5}, which we will sometimes call the {\it family gap bound}, is a natural relaxation of \eqr{v1}: it tolerates sparse or intermittent behavior in either $\Dgap(N)$ or $\dc_N$ while imposing a cumulative bound in the approach to large $N$. See Figure \ref{v1v5fig}. As such, its closure properties, given below, also neutralize many artificial parametrization objections to \eqr{v1}. Note that, by summation, \eqr{v1} implies \eqr{v5}. Conversely, by positivity and reality of $\Dgap$ in unitary CFTs, \eqr{v5} implies a pointwise bound at large $N$:\foot{We sometimes write the sum over $\cT_M \in \cF_N$ simply as a sum over $M$.}
\e{v5end}{\Dgap^d(N) \leq \sum_{M=1}^N \Dgap^d(M) = O(c_N)}
This replaces the central charge increment $\dc_N$ of \eqr{v5} by $c_N$, and is thus strictly weaker than \eqr{v5}. Nevertheless, \eqr{v5end} will itself have strong implications for holography.

\ssec{Power law scaling}

Consider the case, familiar from holography, in which $\Dgap$ and $c_N$ exhibit power law scaling at sufficiently large $N$. We parameterize these as
\e{p1}{c_N \sim N^z\,,\quad \Dgap \sim N^\a}
as $N\rar\i$, with positive $O(1)$ coefficients implied and $z\geq 1$ (cf. \eqr{eq:add-liminf}). Then $\<\Psi\>_\cF<\i$ implies
\e{expbound}{d\a+1\leq z}
We will analyze the holographic implications of \eqr{expbound} in the next section.

Note that the inequality \eqr{expbound} becomes strictly weaker under thinning of the family, which effectively increases $z$. For instance, if $N(M) \sim M^p$ for a thinning exponent $p>1$, then summing over the ordinal $M$ of the thinned sequence would give $p\a d+1 \leq pz$, which is obviously implied by \eqr{expbound}. 

Note also that the same bound would also follow from the unaveraged bound $\Psi<\i$, were we to additionally assume that the sequence $\{c_N\}$ has sufficient regularity that\foot{Note that this does not follow from $c_N\sim N^z$, because it requires extra step-wise regularity along the path.}
\e{p2}{\dc_N \sim N^{z-1}\,,}
but we emphasize that \eqr{expbound} follows from the weaker, averaged bound under \eqr{p1}.

\ssec{Self-averaging in $N$}

Under what conditions does 
\e{selfav}{\<\Psi\>_\cF <\i \quad \Longrightarrow \quad \Psi<\i\,?}
More fundamentally, {\it when is $\Psi$ self-averaging over the family $\cF$, such that $\<\Psi\>_\cF \sim \Psi$?} 

The family-level bound on $\<\Psi\>_\cF$ controls the higher-spin accumulation in central charge units, while the local bound on $\Psi$ controls individual steps in the asymptotic approach. In order for these quantities to be parametrically equivalent, the changes $\cW_+$ must exhibit sufficient regularity that forbids both numerator spikes and denominator dips. 

It is straightforward to formalize sufficient conditions as follows. Consider the dyadic interval $I(N) := [N,2N]$ of length $N$ as $N\rar\i$. Suppose there exist constants $A<\infty$, $B<\infty$, $N_0<\i$, all independent of $N$, such that for every $N\geq N_0$, $\Dgap(N)$ and $\dc_N$ obey 
\e{eq:no-spike}{\Dgap^d(N)\leq \frac{A}{N}\sum_{M=N}^{2N}\Dgap^d(M)}
and
\e{eq:no-dip}{\delta c_N \geq {B\over N} \sum_{M=N}^{2N} \delta c_M = B{c_{2N}-c_{N-1}\over N} \gtrsim {c_{2N}\over N}}
Then $\<\Psi\>_\cF \sim \Psi$. 

Condition \eqref{eq:no-spike} is a no-spike condition: the gap weight at $N$ is parametrically upper-bounded by its average over $I(N)$. Condition \eqref{eq:no-dip} is a no-dip condition: the incoming central-charge increment at $N$ is parametrically lower-bounded by its average over $I(N)$. There are several behaviors of $\Dgap(N)$ and $c_N$ which satisfy these conditions.\foot{See Appendix \ref{appc} for simple artificial examples of intermittent behavior that violate these conditions.} We treat them in decreasing order of strength:
\begin{itemize}
\item If $\cF$ realizes the full set of power law scalings \eqr{p1} and \eqr{p2} at sufficiently large $N$, it is clear that both conditions above are met:
\e{}{\{c_N \sim N^z\,,~ \dc_N \sim N^{z-1}\,,~ \Dgap(N) \sim N^\a\}\quad\Longrightarrow\quad  \<\Psi\>_\cF \sim \Psi}

\item If $c_N$ and $\dc_N$ have the above power-law scalings, but $\Dgap(N)$ is only eventually non-decreasing, then every term in the sum of \eqr{eq:no-spike} is lower-bounded by $\Dgap^d(N)$. Thus, we also have the weaker implication 
\e{}{\{c_N \sim N^z\,,~ \dc_N \sim N^{z-1}\,,~ \text{$\Dgap(N)$ non-decreasing} \}\quad\Longrightarrow\quad  \<\Psi\>_\cF \sim \Psi}

\item We can weaken the central charge requirement in the language of smooth interpolation. Suppose $c(x)$ interpolates $c_N$, and there exist $N$-independent constants $A>0,B>0$ such that, at sufficiently large $N$,
\e{}{\delta c_N\geq A\,c'(N) \geq B{c_{2N}\o N}\,.}
This is more general than the power law package, as it allows for logarithmic corrections and does not even require regularity in $N$. Then together with $\Dgap(N)$ non-decreasing,
\e{}{\{\text{$c_N$ smooth interpolation}\,,~ \text{$\Dgap(N)$ non-decreasing} \}\quad\Longrightarrow\quad  \<\Psi\>_\cF \sim \Psi}
\end{itemize}
As always, these implications are to be understood as holding eventually in $N$. 

In short, and essentially: if $\dc_N$ and $\Dgap(N)$ take their {\it average values} over an asymptotic interval, then $\<\Psi\>_\cF\sim \Psi$, and therefore the self-averaging property \eqr{selfav} holds.\foot{Self-averaging can be rephrased as a property of a uniformly sampled member: within some fixed band at sufficiently large $N$, $\dc_N$ and $\Dgap(N)$ take their average values in the band.} Given that the power law behavior is realized in every known example of large $N$ families, it is natural to wonder whether this self-averaging property -- which follows from strictly weaker assumptions yet -- holds for all families $\cF$ with large gap, as some consequence of special structure of Einstein gravity. 

\ssec{Closure properties}
\label{app:closure}

The family gap bound \eqr{v5} exhibits useful closure properties. These are useful to state in addressing ad hoc constructions, built from combining known natural families of CFTs.\foot{We thank Xi Dong for questions on this point.} 

\sssec{Thinning}

Thinning only helps to satisfy the inequalities. Let $\cF_{\g(n)} := \{\cT_{N(n)}\} \subset \cF$ be an infinite subfamily, with $n_1<n_2<\cdots$. The family is $c$-ordered. The central charge increment thus obeys, at sufficiently large $n$,
\e{eq:thinning-v1-denominator}{\widetilde{\delta c}_{N(n)} :=c_{N(n)}-c_{N(n-1)} =\sum_{M=N(n-1)+1}^{N(n)}\delta c_M \geq\delta c_{N(n)}.}
Consequently, the thinned ratio, call it $\widetilde \Psi(n)$, is upper-bounded by the original ratio $\Psi(N(n))$,
\e{eq:thinning-v1}{\widetilde\Psi(n) := {\Dgap^d(N(n)) \o \widetilde{\dc}_{N(n)}} \leq \frac{\Dgap^d(N(n))}{\delta c_{N(n)}}
=\Psi(N(n)).}
So as $n\rar\i$,
\e{}{\Psi<\i \quad \Longrightarrow \quad \limsup_{n\rar\i}\widetilde\Psi(n)<\i \quad \Longrightarrow \quad \limsup_{n\rar\i}\<\widetilde\Psi\>_{\cF_{\g(n)}}<\i}
Both \eqr{v1} and \eqr{v5} thus survive arbitrary thinning to any admissible subfamily $\cF_{\g(n)} \subseteq \cF$. In particular, the composites of the changes omitted from the thinned subfamily $\cF_{\g(n)}$ belong to $\Wp$, so the thinning always remains admissible. 

\sssec{Finite mixing}

Take finitely many families each obeying \eqr{v5}, and form a $c$-ordered selection from their union. Call the set of families $\{\cF^{(a)}\}$ with $a=1,2,\ldots,a_*<\i$, and call the resulting merged family $\mathsf{F}$. This operation may insert terms between adjacent members of any one parent family $\cF^{(a)}$. If there remains a group of bona fide positive changes $\cW_+$ between the members of the merged family at sufficiently large $N$, then one could consider $\mathsf{F}$ a family in the sense defined above. However, even if not, \eqr{v5} will continue to hold as an inequality so long as $\mathsf{F}$ still respects the {\it ordering condition} \eqr{orderdef} required for \eqr{v5}. So,
\e{}{
\<\Psi\>_\mathsf{F}<\i
\quad\Longleftrightarrow\quad
\bigwedge_{a=1}^{a_*}\<\Psi\>_{\cF^{(a)}}<\i
\label{eq:finite-mix-equivalence}
}
The $\Longrightarrow$ is the statement of thinning given above (which holds in general), and the $\Longleftarrow$ is the statement of mixing (which requires the ordering proviso).\foot{For an example that does not work, consider the merger of two families whose central charge scalings have a $1/N$ splitting at large $N$. This family would not be strictly $c$-ordered, because $\dc_N$ vanishes asymptotically along a subsequence.} In this sense, an ``unnatural'' mixture, such as interleaving a family of free theories within an interacting family every now and then, can still satisfy the bound.  

\sec{Holographic implications}\label{sec6}

We now extract the implications of \eqr{v1}, \eqr{v5} and \eqr{v5end} for holography. The perspective of this section is to grant the family gap bound, and ask what it would imply in general for AdS quantum gravity duals to CFT families $\cF$ which satisfy it.

It is a celebrated conjecture \c{HPPS}, for which there is copious evidence \c{FitzpatrickKaplanAdSEFT,HartmanJainKundu, AfkhamiJeddiHartmanKunduTajdini, MeltzerPerlmutter, Cordova:2017zej, sharp}, that $\Dgap$ sets the holographic cutoff scale of AdS$_{d+1}$ gravitational effective field theory (EFT) in some dimension $D \geq d+1$. We adopt this standard perspective here. In the bulk-boundary map, $\Dgap$ maps to the mass scale of elementary higher-spin fields in the bulk, $M_{\rm HS}$. 

In this section, we drop the explicit $N$ labels on the central charge and the higher-spin gap,  working at asymptotically large $N$. We do not make any assumptions on the $N$-dependence  of these data until otherwise noted. When power law scaling of $c$ and $\Dgap$ is assumed, their exponents are $z$ and $\a$, respectively, as in \eqr{power}. We set $L_{\ads}=1$. 

\ssec{No Planckian scale separation}\label{s61}

Even the weakest form of these bounds rules out AdS scale separation to the Planck scale. By the latter, we mean a gravitational EFT that remains parametrically valid up to the $(d+1)$-dimensional Planck scale,
\e{}{M_{p,d+1} \sim c^{1/(d-1)}}
AdS scale separation is commonly stated in geometric terms (``no large extra dimensions''), but regardless of whether an AdS vacuum admits a geometric description, an agnostic precondition for the validity of the semiclassical solution is that the EFT remains valid up to $M_{p,d+1}$: in other words, it would not be meaningful to claim parametric scale separation beyond a scale at which the EFT has already failed. We emphasize that the question of {\it parametric} scale separation is necessarily a {\it family-level} question: it concerns the behavior of families of AdS/CFT pairs in the approach to large $N$.\foot{One may also ask about sporadic quantum gravity vacua exhibiting large but finite hierarchies, as in the AdS vacua of KKLT type (e.g. \c{KKLT,Demirtas:2021nlu,Demirtas:2021ote}). This is a different question, and we have nothing to say about it.} As such, a family-level constraint seems necessary to properly analyze it. 

It is immediate to see that \eqr{v5end} forbids Planckian scale separation. Setting $\Dgap \sim c^{1/(d-1)}$, we observe
\e{}{\Dgap^d \sim c^{d/(d-1)} \gg c\,.}
This violates \eqr{v5end} for all $d>1$. 

More generally, \eqr{v5end} implies a rate bound on the ratio of higher-spin and Planck scales: 
\e{hspend}{ \frac{M_{\rm HS}}{M_{p,d+1}}\sim \frac{\Delta_{\rm gap}}{c^{1/(d-1)}} \lesssim c^{-1/(d(d-1))}}
which vanishes in the large $c$ limit. An equivalent formulation is that the dimensionless gravitational coupling at the higher-spin/EFT scale, call it $g^2_{\rm grav}(\mhs)$, is suppressed as
\e{eq:grav-coupling}{g^2_{\rm grav}(\mhs) := G_{d+1}M_{\rm HS}^{d-1} \sim \frac{\Dgap^{d-1}}{c}\lesssim{1\o\Dgap}}
Gravity is therefore parametrically weak at the scale where the EFT breaks down, by at least one inverse power of the gap.\foot{Note that the weakly coupled regime is precisely where the bootstrap bounds of \c{sharp} on EFT couplings are valid.} In the next subsection we will derive a stronger version of this using the full family gap bound \eqr{v5}. 

Is there a tower? The above implication says nothing about the spectrum of low-spin ($J\leq 2$) states or KK towers. On the other hand, if one assumes {\it convexity} of the leading single-trace Regge trajectory in the $\D$-$J$ plane, then there is an entire tower of single-trace states with spins $J>2$ and conformal dimensions $\Delta_J=O(\Dgap)$. In other words, assuming single-trace convexity, Planckian scale separation {\it is} ruled out by the presence of a tower, in this case, a higher-spin tower. While convexity of the leading non-perturbative (finite $N$) trajectory has been proven rigorously in CFT \c{CHP}, this has not been extended to single-trace trajectories at large $N$, though it has often been assumed and observed to be the case.

An extreme case of Planckian scale separation is a theory of ``pure gravity'', in which there are {\it no} states up to a threshold $\D \sim M_{p,d+1}$. In AdS$_3$, the bound \eqr{v5end} implies $\Dgap = O(\sqrt{c})$, and would thus rule out AdS$_3$ pure gravity quite strongly, understood in the sense of a semiclassical limit of a family $\cF$ of the type described here. Ordinarily, attempts to bootstrap or rule out pure gravity do not focus on the higher-spin sector. Note that this is not in conflict with recent work viewing AdS$_3$ pure gravity as computing averages of CFT data, because that correspondence is not formulated for the full gravitational path integral. The implication for other bulk constructions with $\Dgap = O(c)$ \c{BCM,DiUbaldo:2023hkc} is that they must not actually arise as semiclassical limits of families. 

\ssec{Bounds on scale separation and extra dimensions}

The family average bound $\<\Psi\>_\cF<\i$ in \eqr{v5} imposes much more stringent constraints than what we have extracted above from its weaker consequence \eqr{v5end}. We are interested in the case where the bulk is a gravitational EFT in dimension $D=d+1+n$, for some number $n\in\Z_+$ of large (i.e. AdS-sized) extra dimensions. Whereas in a ``pure'' gravitational EFT in $D$ dimensions, $\Dgap \sim M_{p,D}$, this may not hold in general, as in weakly coupled string theory where the ratio is controlled by a power of $g_s\ll1$. We will derive bounds on the ratio $M_{\rm HS}/M_{p,D}$ that follow from \eqr{v5}.

We first relate $c$ to $M_{p,D}$ in the standard way. For concreteness and intuition, we phrase things geometrically, though the generalization will be obvious enough in what follows. For an Einstein-frame metric of warped form,
\e{}{ ds_D^2=e^{2A(y)}ds_{\mathrm{AdS}_{d+1}}^2+g_{ij}(y)\,dy^i dy^j}
where $i=1,\ldots,n$ runs over dimensions of a manifold $X_n$, dimensional reduction gives the AdS$_{d+1}$ Newton's constant,
\e{GGV}{\frac{1}{G_{d+1}} =\frac{\Vol(X_n)}{G_D} \,,\quad\text{where}\quad \Vol(X_n) := \int d^ny\,\sqrt{g(y)}\,e^{(d-1)A(y)}}
The holographic dictionary is thus
\e{}{c \sim \Vol(X_n) (M_{p,D})^{d-1+n}}
In what follows we take $\Vol(X_n) = O(1)$ uniformly in $N$ --  a fixed, nondegenerate internal geometry and a bounded warp profile at sufficiently large $N$ are sufficient.\footnote{If the volume parametrics vary non-trivially with $N$ as one approaches $N\rar\i$, the bounds to follow acquire a simple volume factor that is easy to track; alternatively, one can interpret them in a limsup sense. Also, let us here make the usual remark that to literally interpret $n$ as a number of {\it large} extra dimensions requires a corresponding isotropy of internal length scales on $X_n$. We use the language of large dimensions henceforth.} In non-geometric setups, $\Vol(X_n)$ is to be interpreted operationally as defining the hierarchy in \eqr{GGV}. 

Now, recall that assuming eventual monotonicity of $\Dgap(N)$, \eqr{v5} implies
\begin{equation}
 \Dgap^d(N)\lesssim \frac{c_N}{N}.
 \label{eq:holo:pointwise}
\end{equation}
Combining the above equations yields the following bound:
\e{}{{M_{\rm HS} \o M_{p,d+1+n}} \lesssim N^{-1/d} c^{(n-1)/(d(d-1+n))}}
For power law scaling $c\sim N^z$, we thus arrive at the following implication of \eqr{v5}:
\e{gpower}{\boxed{\frac{M_{\rm HS}}{M_{p,d+1+n}}\lesssim N^{\g(d,n,z)}\,,\quad \text{where}\quad \g(d,n,z) := \frac{(n-1)(z-1)-d}{d(d-1+n)}}}
This is one of our central holographic results. 

One immediately sees two distinct uses of \eqref{gpower}.

First, suppose that the bulk EFT is valid up to the $D$-dimensional Planck scale,
\e{}{M_{\rm HS}\sim M_{p,d+1+n}\,,}
as in strongly coupled string/M-theory. This is possible if and only if
\e{ggeq0}{\g(d,n,z)\geq 0}
In other words, given a central charge scaling exponent $z$ for a family $\cF$ of $d$-dimensional CFTs, the number of large extra dimensions is {\it lower-bounded},  
\e{nbound}{{n \geq 1 + {d\o z-1}}}
This does not use R-symmetry or string/M-theory. 

As an example, a four-dimensional matrix-like family $\cF$ with $z=2$ requires at least five AdS-sized internal dimensions under these assumptions. This accounts for the property of known supersymmetric AdS$_5 \x SE_5$ Freund-Rubin families in string theory, whose five large extra dimensions has always defied fundamental explanation: in particular, even granting the geometrization of R-symmetry (itself not a proven fact about holography), the presence of {\it five} extra dimensions exceeds what sub-maximal supersymmetry would seem to require. 

If, on the other hand, $n$ is {\it smaller} than the above threshold -- i.e. if $\g(d,n,z)<0$ -- then the higher-spin scale must separate parametrically from $M_{p,d+1+n}$ at a bounded rate in the semiclassical limit $G \rar 0$:

\begin{itemize}

\item At $n=0$, 
\e{}{\g(d,0,z) = -{d-1+z\over d(d-1)}<0}
This strengthens the prohibition of Planckian scale separation by providing a maximal ratio,
\e{}{\frac{M_{\rm HS}}{M_{p,d+1}} \lesssim c^{\g(d,0,z)/z} \quad \Longleftrightarrow \quad g^2_{\rm grav}(\mhs) \lesssim c^{-(d-1+z)/dz}}
which constrains how slowly a proposed weak-coupling limit may approach its boundary.\foot{In terms of $\Dgap$, $g^2_{\rm grav}(\mhs) \lesssim\Dgap^{-\(1+{1/\a}\)}$.} Note that the ratio is suppressed by an extra $c^{-1/(dz)}$ relative to the endpoint bound \eqr{hspend}, because we used the full family average structure.
\vs
In other words, this is a constraint on {\it stringy scale separation}, reformulated agnostically in terms of $\mhs$ instead of $M_s$: so long as the $(d+1)$-dimensional theory is sufficiently weakly coupled, scale separation is possible. 
\vs
\item At $n=1$, 
\e{}{\g(d,1,z) = -{1\over d}<0}
This is an especially clean consequence: a single large extra dimension would require a hierarchy with a minimum decay rate in $N$ that is {\it independent of $z$}. 
\vs
We can also turn the logic around by backing off from the power law growth ansatz, and asking what growth of $c$ {\it is} compatible with Planckian EFT with $n=1$. It is straightforward to show that \eqr{v5}, with no further regularity or scaling assumptions, implies 
\e{}{n=1 \quad \Longrightarrow \quad c_N \gtrsim e^{\kappa N}}
for some $\kappa>0$. (We remind the reader that there is no known controlled example of an AdS/CFT dual pair with $c_N$ growing super-polynomially in a physical parameter $N$.)
\vs
\item At $n\geq 2$, a Planckian EFT in AdS$_{d+1}$ {\it is} possible for a power-law family, provided that the central charge scales fast enough as $N\rar\i$. For $n=2$,  
\e{}{\g(d,2,z) = {z-(d+1)\o d(d+1)} \geq 0 \quad \Longleftrightarrow \quad z \geq d+1}
This growth is saturated by the MSW CFT, with triplet $(d,n,z)=(2,2,3)$ -- precisely as it must be for a CFT dual to an AdS$_3 \x S^2 \x CY_3$ background of M-theory. 

\end{itemize}

In the case of $d=4$ CFT families with $z=2$, mentioned earlier, such a family could have an AdS$_5$ string dual with $n<5$ only if the five-dimensional string coupling is driven to zero (unlike the Freund-Rubin case).

\subsubsection{Comment on DGKT and LVS}

As is visible in Figure \ref{survey}, the DGKT solution obeys \eqr{v1}. The standard scaling is, with four-form flux number $N$, $c_N\sim N^{9/2}$ and $\ell_s^{-1} \sim N^{3/4}$, with vanishing string coupling $g_s\asymp N^{-3/4}$. Thus, if we presume that the putative $d=3$ CFT is a regular family for which $\dc_N \sim N^{7/2}$, taking $\Dgap \lesssim \ell_s^{-1}$ gives
\e{}{\Psi = O(N^{-5/4}) = O(g_s N^{-1/2})}
which satisfies \eqr{v1} easily. In the brane interpretation, the relevant charge-changing D4 domain wall wraps a two-cycle whose string-frame area grows as $v\sim N^{1/2}$; so the residual negative power of $N$ is therefore a physical wrapped volume, in conjunction with our earlier discussion. So far as we have checked, other more recent DGKT-type constructions in $d=2,3$ also obey the bound. 

The Large Volume Scenario \cite{LVS, LVS-CFT} presents a further challenge, because it is even less clear that the dual CFT should furnish a family, or how to identify $N$. Nevertheless, using standard relations seems to give consistency with \eqr{v1}. We defer a proper analysis.

\ssec{Consequences for string/M-theory}
Noting that 
\e{}{{\partial\over \partial n} \g(d,n,z) = {z\over (d-1+n)^2}>0\,,}
we can upper bound $\g(d,n,z)$ in string/M-theory for any $n$ by setting $D=10,11$. Writing $n=9-d+\vareps$, 
\e{}{\g_{\rm string/M} \leq \g(d,9-d+\vareps,z) = {z-1\over d} -{z\over 8+\vareps}\,, \qquad \vareps = \begin{cases}
0&~~\text{(string theory)}\\
1&~~\text{(M-theory)}
\end{cases}}
Now, as in \eqr{ggeq0}, if there is a genuine M-theory background (i.e. not dualizable to a weakly-coupled string) with an AdS$_{d+1}$ factor, the upper bound must be positive. This implies a lower bound on $z$:
\e{stringMdim}{{z \geq {8+\vareps\o 8+\vareps-d}}}
This is a string/M-theory upgrade of the previous results: when the spacetime dimension $n$ cannot be too large, $z$ cannot be too small. In particular, 
\es{}{\forall~\text{AdS$_4$ M-theory vacua:}& \quad z\geq {3/2}\\
\forall~\text{AdS$_5$ string vacua:}& \quad z\geq 2\\
\forall~\text{AdS$_7$ M-theory vacua:}& \quad z\geq 3}
and so on. This holds for AdS$_{d+1} \x X_n \x \text{(small)}$ backgrounds for any $n$. 

This may be viewed as a type of explanation of the origin of the $c \sim N^3$ scaling in the $(2,0)$ CFT.\foot{A convincing explanation is already known \c{CDY}.} Let us make precise what we mean, as a statement in field theory. Letting $\cF$ denote the complete definition of families for which \eqr{v5} holds, its implied bound \eqr{stringMdim} identifies the $N^3$ scaling as the slowest scaling possible for a $d=6$ CFT $\cT\in\cF$ with an M-theory dual, and as impossible with a string theory dual, which would require at least $N^4$. This can be stated in purely field-theoretic terms, because the restriction to $D\leq 11$ effective bulk dimensions can be derived, in principle, using the subleading corrections to anomalous dimensions of double-trace composite operators \c{AP}. Thus, from this perspective, the problem of giving a fundamental field-theoretic explanation of the $N^3$ scaling reduces to showing that the $(2,0)$ CFT belongs to $\cF$.

\ssec{Extremal families}\label{s64}
It is natural to ask and answer the following two questions in this context:

\begin{enumerate}[label*=  \bf \arabic*)]
\item {\it Which backgrounds saturate the family gap bound?} For power law families,  $d\a+1=z$. 

\item {\it Which backgrounds are vacua of a Planckian EFT?} For power law families, $z=\a(d-1+n)$.  
\end{enumerate}

Let us call a power law family ``extremal'' if it saturates both of these scales: that is, an {\it extremal family} has the property that in the bulk semiclassical limit, the higher-spin scale is maximized with respect to both the gap bound \eqr{v5} and the Planck scale of gravitational EFT in dimension $D=d+1+n$ (for some $n$). Putting these together, we deduce the following property of extremal families:
\e{ext}{\textbf{Extremal families:} \qquad (z,\a) = \(1+{d\o n-1},{1\o n-1}\)}
This property is obeyed by $\cN=4$ SYM, ABJM for all $k\lesssim N^{1/5}$, the (2,0) CFT, the D1--D5 system in the matrix limit $N_1\sim N_5 \sim N$, the MSW CFT in the cubic limit $p_1,p_2,p_3 \sim N$, and orbifolds thereof. Indeed, all Freund-Rubin families enjoy this property, thanks to the flux quantization condition $N \sim (M_{p,D})^{n-1}$. 

The integrality of dimensions $d$ and $n$ implies that in any extremal family, the scaling exponents of $c$ and $\Dgap$ are {\it rational}:
\e{}{d,n\in \Z_+ \quad \Longrightarrow \quad z,\a \in \mathbb{Q}\,.}
This is a constraint on the semiclassical scaling of extremal families of AdS vacua in any theory of quantum gravity, again realized nicely in string/M-theory.

\sssec{Explaining $z=d/2$}

Now, suppose we further set $z=d/2$. Plugging into the above we find
\e{}{z={d\o 2} \quad\Longrightarrow \quad n = 1+{2d\o d-2}\qquad}
The interesting result is that $d=3,4,6$ are the {\it only} solutions with $n \in \Z_+$. It is intriguing that integrality of spacetime dimension gives a constraint that singles out the canonical conformal branes of string/M-theory.

To explain the pattern more satisfactorily, though, we should really explain the converse property, i.e. in what sense $z=d/2$ is singled out; after all, any $(d,n)$ gives an allowed doublet $(z,\a)$ in \eqr{ext}. To do so, we restrict to critical string and M-theory, by plugging in $n=9-d$ or $10-d$, respectively. The result is that the $d=3,4,6$ solutions are the {\it unique solutions with $z=d/2$}. 

Thus, to summarize, the $z=d/2$ pattern of the D3-, M2- and M5-brane backgrounds is a certain {\it extremization property of string/M-theory}: it is the result of maximizing the family gap bound in a purely gravitational EFT in the critical spacetime dimension. Conversely, these are the unique extremal family scalings with $z=d/2$ for which $d\in \Z, n\in\Z_+$. The ``truly'' fundamental pattern is not so much $z=d/2$, but the general extremization condition \eqr{ext} defining extremal families; this extends the extremal set to include the canonical two-dimensional case of the D1--D5 CFT (as well as the MSW CFT, for good measure), which provides a satisfying unification of the canonical constructions. 

\ssec{Pure AdS$_5$ and exotic $\cN=4$ SCFTs}

The case of $d=4$, maximally-supersymmetric CFTs deserves special treatment. It is an open problem to determine whether there exist ``exotic'' -- that is to say, non-SYM -- $\cN=4$ SCFTs in four dimensions \c{holo:BRR}. The large $N$ version of this question, interpreted holographically, is whether there exist classical, maximally-supersymmetric AdS$_5$ vacua which are not 
the AdS$_5 \x S^5$ solution of type IIB string theory. Within this exist variations: one can demand different degrees of scale separation (e.g. Planckian vs. stringy), different degrees of completeness of the half-BPS superconformal primary spectrum, eschew string/M-theory altogether, or other intermediate possibilities. For previous arguments, see \c{holo:AldayChester,holo:MRV}.

Consider a putative family of exotic $\cN=4$ theories with a large gap. Let us ask what the family gap bound implies. We have already shown, in Section \ref{s61}, that Planckian scale separation (``AdS$_5$ pure maximal supergravity'') is ruled out by this bound. However, the $\cN=4$ case has an extra feature: by superconformal representation theory, every local $d=4$, $\cN=4$ SCFT has a complexified exactly marginal coupling, call it $\t$, which couples to the stress tensor multiplet. Because there is a unique $\cN=4$ Lagrangian, namely, that of the SYM theory, with coupling constant $g_{\rm YM}^2 \sim \Im(\t)^{-1}$, any exotic $\cN=4$ SCFT must have a conformal manifold which excises the free cusp at $\Im(\t)\rar\i$.  

This extra feature allows us to rule out even {\it stringy} scale separation, as we now show. From \eqr{gpower}, the gap bound is
\e{}{\({M_s\o M_{p,5}}\right)^3 \lesssim N^{-{(z+3)/4}}\,.}
Suppose that there is a type II string realization with $\mhs \sim M_s$ and a well-defined physical coupling, $g_s$, that we identify with $\Im(\t)^{-1}$ in the CFT. We define the internal normalization $V_s$ operationally by
\e{}{\({M_s\o M_{p,5}}\right)^3 \sim {g_s^2\o V_s}\,.}
For a geometric compactification, $V_s$ is the internal volume in string units (if we write AdS$_5 \x X_5$, then $V_s := \Vol(X_5)M_s^5$, where $\Vol(X_5)$ is the volume in AdS units). Combining these two relations, once we demand that our exotic family obey $g_s>0$, that implies 
\e{}{V_s \rar\i\,.}
This rules out stringy scale separation in the semiclassical limit: at least some component of the internal space must diverge relative to the string scale.\foot{Again, we have used geometric language for clarity, but the conclusion would equally apply in a non-geometric compactification with $V_s$ interpreted as setting the five- versus ten-dimensional gravitational couplings.} Passing to AdS units using $\mhs \sim M_s$ and standard relations yields a further refinement:
\e{}{\Vol(X_5) \gtrsim g_s^2 N^{2-z}\,.}
Therefore, if we insist on $n=0$, which is the condition $\Vol(X_5) \ll 1$, then $z>2$. 

This analysis strongly corners the space of possible families $\cF$ of exotic $\cN=4$ SCFTs at large $N$ that the gap bound \eqr{v5} would allow: they must exhibit a faster-than-$N^2$ growth of degrees of freedom, with a bulk dual exhibiting sub-string scale hierarchies or, perhaps, not being a string theory at all.

\ssec{Further swampland and spectral connections}
The preceding results may all be read as swampland constraints in AdS. Let us also make a few further remarks on relations to other swampland literature. 

\sssec{Relation to the species scale}\label{sec:holo:species}

The species scale $\Lambda_{\rm sp}$ is often treated as the scale up to which a gravitational EFT remains valid \c{Dvali:2007hz}.\foot{The actual relation between the species scale and what one should identify as the genuine EFT cutoff is not universal. Pragmatically, precise cutoff prescriptions may differ at the level of logarithmic corrections. For a critical analysis of the species scale, and an alternative proposal for what the EFT scale should be, see \cite{CHL}. For an overview of recent work in this area, see \c{Herraez:2025clp}. Our analysis in this subsection is not meant to adjudicate these matters but simply to say what the family gap bounds imply for species counts as conventionally defined. As an aside, the higher-spin onset scale, $\Lambda_{\rm o}$, of \cite{CHL} seems potentially useful in seeking a proof of \eqr{v5} or an OPE-weighted version thereof.} More precisely, it is the energy at which the graviton propagator is no longer dominated by its tree-level approximation. In $D$ dimensions, in a particle regime its parametric relation to the number of effectively light species is
\begin{equation}
 N_{\rm sp}\sim
 \left(\frac{M_{p,D}}{\Lambda_{\rm sp}}\right)^{D-2}.
 \label{eq:holo:speciesdefinition}
\end{equation}
It is a separate physical assumption, well-motivated in AdS, that $\Lambda_{\rm sp}\sim M_{\rm HS}$. When this holds (possibly modulo logs), we can trivially recast the scale separation bound \eqr{gpower} as a bound on the number of species. Assuming power law scaling $c\sim N^z$ for simplicity and recalling $D=d+1+n$,
\begin{equation}
 N_{\rm sp}\gtrsim N^{\,1-(n-1)(z-1)/d}.
 \label{eq:holo:speciesrate}
\end{equation} 
The estimate is informative as a growth bound when its exponent is
positive. In particular, in the presence of zero or one large extra dimension, 
\es{}{n=0:&\quad N_{\rm sp}\gtrsim N^{1+{z-1\o d}}\\
n=1:&\quad N_{\rm sp}\gtrsim N.}
We remind the reader that $N$ is the ordinal coordinate of the CFT family $\cF$ whose dual AdS vacua we are considering, a priori not identified with flux or other top-down bulk quantities.

\sssec{Relation to the Emergent String Conjecture}\label{sec:holo:ESC}

The Emergent String Conjecture (ESC) \c{holo:ESC} asserts that infinite-distance limits on moduli space must either be associated to spacetime decompactification or an emergent weakly coupled string description \cite{holo:ESC}. Equation \eqr{gpower} has a loosely related interpretation at the level of a family of quantum gravity theories, in the following sense: if that ratio is bounded away from zero, the dimension must satisfy \eqr{nbound} -- this is decompactification; if not, the inequality quantifies the hierarchy required for the chosen $n$ -- this is the abstraction of a weakly coupled string.

Of course, a defining point about the ESC is that these weakly coupled limits must {\it actually} be strings per se. The analogy above is weaker, solely one of scope: \eqr{gpower} gives a family-level analog of the {\it hierarchy aspect} of the ESC, with bounds on rates of change along a discrete ordinal family. In contrast to the ESC, \eqr{gpower} operates in theory space: it does not require a moduli space. We note that existing species and distance conjectures already constrain rates with respect to field-space distance \cite{holo:SpeciesDistance}. Here the parameter is the number of selected theories in an admissible family $\cF$ or subfamily thereof. 

\sssec{Counting AdS vacua}

We consider to what extent our family-level bound refines the counting of AdS vacua of \c{BTV}. Their normalized cutoff $\hat \mu$, measured in $(d+1)$-dimensional Planck units, is\foot{Note that our $d+1$ is their $d$.}
\e{}{\hat\mu=\frac{\mu}{M_{p,d+1}}\,,}
where $\mu$ is the gravitational EFT cutoff. Using our identification of $M_{\rm HS}$ as (an upper limit on) the cutoff,  
\e{}{\hat\mu
\lesssim
\frac{M_{\rm HS}}{M_{p,d+1}}
\sim
\frac{\Delta_{\rm gap}(N)}{c_N^{1/(d-1)}}.}
In our earlier notation \eqr{eq:grav-coupling}, 
\e{}{\hat\mu^{d-1} \lesssim g^2_{\rm grav}(M_{\rm HS}) \lesssim {1\o \Dgap}\,,}
where the first inequality is just definitions, and the second is imposed by the family gap bound. 

In the approach of \c{BTV}, one counts AdS vacua in the limit $\hat \mu \rar 0$. Let $\mathcal N_\cF(\hat\mu)$ denote the number of members of a given family $\cF$ below the cutoff. A general consequence of \eqr{v5}, assuming no extra structure (e.g. power law scaling), is that 
\e{cons}{\mathcal N_\cF(\hat\mu) \lesssim \hat\mu^{-(d-1)} \qquad (\hat\mu\rar0).}
To prove this, we fix $\cF$ and consider a central charge interval $c_M \in I(C) := [C,2C]$. Let $N$ be the ordinal index such that $c_N := 2C$. The sum over theories $\cT_M$ with $c_M \in I(C)$ which also obey the cutoff is a subset of all theories in that range; so \eqr{v5} implies
\e{}{\sum_{\substack{c_M \in I(C)\\ \cT_M\ \mathrm{passes\ cutoff}}}
\Delta^d_{\rm gap}(M) \leq \sum_{M=2}^{N}\Delta^d_{\rm gap}(M) \lesssim 2C.}
Every term obeys two lower bounds on its contribution:
\e{}{\Delta^d_{\rm gap}(M)\gtrsim1, \qquad \Delta^d_{\rm gap}(M) \gtrsim \hat\mu^d c_M^{d/(d-1)} \geq \hat\mu^d C^{d/(d-1)}.}
The first is unitarity, while the second is the definition of the cutoff $\hat\mu$ for $c_M \in I(C)$. Consequently,
\e{}{\#\{M:c_M \in I(C),\ \cT_M\text{ passes cutoff}\} \lesssim \min\!\left\{C,\,\hat\mu^{-d}C^{-1/(d-1)}\right\}.}
The two terms cross at $C \sim \hat \mu^{-(d-1)}$. By dyadic summation, one obtains \eqr{cons}.\footnote{i.e. the bins near the crossover dominate the total count: the bounds grow geometrically before the crossover, and decay geometrically afterwards.} 

For families $\cF$ with regular power law scaling of $c$ and $\Dgap$, this upgrades to a sharper result,
\e{}{\mathcal N_\cF(\hat\mu) \lesssim N =  \hat\mu^{-1/\left(z/(d-1)-\alpha\right)}\qquad (\hat\mu\longrightarrow0)\,.}
The exponent increases monotonically from zero with $\alpha$, so imposing the family bound $d\a +1 \leq z$ gives 
\e{}{{\mathcal N_\cF(\hat\mu) \lesssim \hat\mu^{-d(d-1)/(z+d-1)}}\qquad (\hat\mu\longrightarrow0)\,,}
which is stricter than \eqr{cons}. 

These bounds on $\mathcal N_\cF(\hat\mu)$ are bounds on the cardinality of {\it individual} families $\cF$. In contrast to the top-down approach of \c{BTV}, they may be viewed as (conditional) bounds on bottom-up counts. However, \c{BTV} also sums over families, within some specified bulk universality class, including volume weights associated with continuous scalar directions. It would be interesting to attempt to merge these perspectives -- in particular, the above bounds can be used to stratify a global count in some class, by dividing it into families and bounding the contribution of any given family.

\sssec{A short remark on erraticness}
A hope for future work is that the concept of families and provenance might be used to shed light on the erratic $N$-dependence of black hole microstates and the filtration thereof. One comment in this regard is that the domain walls in our work and in \c{KFW} are both mediated by the same generalized branes, just placed in different backgrounds -- empty AdS here, and the double-cone there. As such, they probe different parts of the spectrum, and it is intriguing to ask whether the respective consequences for spectra may be unified. In other words, perhaps these branes, and the families $\cF$ into which they are embedded, could help explain how AdS locality at the gap scale implies statistical universality at the black hole scale. At a narrower level, we wonder whether our notion of provenance could be a useful precondition for the analytic continuation in $N$ that is the starting point for the averaging procedure of \c{KFW}. 

\sec*{Acknowledgments}

Earlier versions of this work were presented in seminars at QMAP, UC Davis in February 2023, and at the KITP program ``What is String Theory? Weaving Perspectives Together'' in January 2024; we thank the participants, especially Xi Dong, David Gross, Ying-Hsuan Lin, Mukund Rangamani and Xi Yin, for stimulating questions. We are grateful to Alex Belin, Nathan Benjamin, Antoine Bourget, Matilda Delgado, Hong Liu, Mukund Rangamani, Edward Witten and Sasha Zhiboedov for helpful discussions. We acknowledge the use of ChatGPT for assistance in honing arguments, making figures and draft editing. We thank Bic for hardware support. We thank the IHES for hospitality during the course of this work. This work was supported in part by ERC Starting Grant 853507, and in part by grant NSF PHY-2309135 to the Kavli Institute for Theoretical Physics (KITP).

\begin{appendix}

\sec{Survey supplement}\label{appa}
Here we register a few complementary details to our survey in Figure \ref{survey}. 

The central charge $c$ is computed holographically as
\e{}{c \sim \frac{L_{\ads}^{d-1}}{G_{d+1}}}
where $G_{d+1}$ is Newton's constant in AdS$_{d+1}$. In holographic contexts, the actual $\Dgap$ is generally not known exactly. All we need are upper bounds. When independent CFT data are not available, we make the conservative choice to use the corresponding bulk string or Planck scale. In a warped solution with string-frame metric of the form
\e{}{ds_s^2=e^{2A_s(y)}ds_{\ads}^2+ds_{\mathrm{int}}^2\,,}
the conservative quantity is the lightest redshifted onset,
\e{}{\Dgap \lesssim \inf_ye^{A_s(y)}\frac{L_{\ads}}{\ell_s}\,.}

Many backgrounds of string/M-theory admit several choices of relative scaling of the flux parameters. Figure \ref{survey} displays the simplest, most common choices -- for example, the $c \sim N^3$ scaling of the MSW CFT. In all examples we have checked, alternative scalings also satisfy $\Psi<\i$. 

The following is a table of references. We have prioritized their utility in checking the scaling behavior. The most mainstream examples can of course be found in \c{Maldacena, Witten, GKP, MAGOO}. 
\vskip .1 in
\begin{center}
\begingroup
\small
\setlength{\tabcolsep}{5pt}
\renewcommand{\arraystretch}{1.17}
\begin{tabular}{@{}c p{5.1cm} l@{}}
\toprule
& Construction & References\\
\midrule
\SurveyDot{1} & $\mathrm{AdS}_4\times X_7$
  & \cite{ABJM,DMP}\\
\SurveyDot{3} & $\mathrm{AdS}_5\times X_5$
  & \cite{KlebanovWitten,GubserEinstein,Morrison:1998cs}\\
\SurveyDot{6} & $\mathrm{AdS}_7\times S^4/\Gamma$
  & \cite{CDY}\\
\SurveyDot{4} & Constant-$\tau$ F-theory AdS$_5$
  & \cite{FayyazuddinSpalinski,AharonyTachikawa}\\
\SurveyDot{5} & ABJM IIA, fixed $g_s$
  & \cite{ABJM,DMP}\\
\SurveyDot{8} & D1--D5, matrix ray
  & \cite{MaldacenaStrominger}\\
\SurveyDot{7} & D1--D5, vector ray
  & \cite{MaldacenaStrominger,AU}\\
\SurveyDot{19} & MSW, cubic ray
  & \cite{MSW}\\
\SurveyDot{14} & F-theory $\mathrm{AdS}_3$
  & \cite{FTheoryAdS3}\\
\SurveyDot{13} & $T[SU(N)]$
  & \cite{AsselEstesYamazaki}\\
\SurveyDot{18} & Massive IIA $\mathrm{AdS}_4$
  & \cite{GuarinoJafferisVarela}\\
\SurveyDot{9} & Class $\mathcal{S}$
  & \cite{GaiottoDualities,GaiottoMaldacena}\\
\SurveyDot{15} & Massive IIA $\mathrm{AdS}_5$
  & \cite{ApruzziMassiveAdS5,BPT}\\
\SurveyDot{16} & Argyres--Douglas
  & \cite{BahArgyresDouglas,BBMN}\\
\SurveyDot{11} & Massive IIA $\mathrm{AdS}_6$
  & \cite{BrandhuberOz,JafferisPufu}\\
\SurveyDot{12} & IIB $\mathrm{AdS}_6$ brane webs
  & \cite{DHokerGutperleUhlemann,FluderUhlemann}\\
\SurveyDot{20} & DGKT
  & \cite{DGKT,Junghans}\\
\SurveyDot{22} & DGKT-like $\mathrm{AdS}_3$
  & \cite{FarakosTringasVanRiet}\\
\bottomrule
\end{tabular}
\endgroup
\end{center}

\sec{Artificial examples with $\<\Psi\>_\cF<\i$ but $\Psi=\i$}\label{appc}

The following simple examples demonstrate the mechanisms behind the possible failure of the step-wise bound $\Psi<\i$, even when the family gap bound $\<\Psi\>_\cF<\i$ holds. The indicated behaviors need only hold eventually in $N$.

{\it Numerator-spike failure:} Take a regular denominator $c_N\sim N^z$ with $z\geq1$ and let
\begin{equation}
\Dgap^d(N)\sim
\begin{cases}
N^z,&N\text{ is a power of }2,\\
1,&\text{otherwise}.
\end{cases}
\label{eq:numerator-spikes}
\end{equation}
Then \eqr{v5} holds, while $\Dgap^d(N)/\d c_N \sim N$ is unbounded along powers of two. 

{\it Denominator-dip failure:} Conversely, take a regular non-decreasing numerator
\begin{equation}
\Dgap^d(N)\sim N
\label{eq:regular-numerator}
\end{equation}
and start from the increments $2N-1$ of $c_N = N^2$. For $n_k:=2^k$, $k\geq2$, modify only two consecutive increments by
\begin{equation}
\begin{aligned}
\delta c_{n_k}&:=n_k^{-2},
\end{aligned}
\label{eq:dip-increment}
\end{equation}
\begin{equation}
\begin{aligned}
\delta c_{n_k+1}&:=2(n_k+1)-1+
\bigl(2n_k-1-n_k^{-2}\bigr),
\end{aligned}
\label{eq:compensating-increment}
\end{equation}
and retain $\delta c_N=2N-1$ otherwise. With $c_1=1$, this gives
\begin{equation}
c_N=
\begin{cases}
(N-1)^2+N^{-2},&N=2^k,\ k\geq2,\\
N^2,&\text{otherwise}
\end{cases}
\label{eq:dip-central-charge}
\end{equation}
The sequence is strictly increasing. \eqr{v5} holds, but at $N=n_k$,
\begin{equation}
\frac{\Dgap^d(n_k)}{\delta c_{n_k}}=n_k^3\longrightarrow\infty\,,
\label{eq:dip-V1-failure}
\end{equation}
violating \eqr{v1}. 

\sec{$q$-ordering does not imply central charge ordering: SQCD}\label{appd}

Consider four-dimensional $\mathcal N=1$ SQCD with gauge group $SU(N_c)$ and $N_f$ fundamental--antifundamental pairs. Inside the conformal window, the exact $a$-type anomaly coefficient is
\e{}{
    a(N_c,N_f)=\frac{3}{16}
    \left(2N_c^2-1-\frac{3N_c^4}{N_f^2}\right),
    \qquad
    \frac32N_c<N_f<3N_c.
}
We fix the electric construction, in which positive charge changes correspond to adding colors and/or flavors. Letting $r:=N_f/N_c$,
\e{}{
    \D a=\frac{3N_c}{4}
    \left(
        \left(1-\frac{3}{r^2}\right)\D N_c
        +\frac{3}{2r^3}\,\D N_f
    \right).
}
Thus a step with $\D N_c,\D N_f>0$ decreases $a$ when
\e{}{
    \frac32<r<\sqrt3,
    \qquad
    \frac{\D N_f}{\D N_c}<\frac{2r(3-r^2)}{3}.
}
Near the lower part of the conformal window, adding colors sufficiently faster than flavors therefore lowers $a$. A conceptually identical analysis applies to $c$ instead of $a$.

The following is an explicit sequence that persists out to infinity. Define
\e{}{
    A_k=2^k(14,25),\qquad
    B_k=2^k(17,26),\qquad k\in\Z_{\geq 0},
}
where each pair denotes $(N_c,N_f)$, and take the path
\e{}{
    A_0\longrightarrow B_0\longrightarrow A_1
    \longrightarrow B_1\longrightarrow A_2\longrightarrow\cdots.
}
Every step strictly increases both charges:
\e{}{
    B_k-A_k=2^k(3,1),
    \qquad
    A_{k+1}-B_k=2^k(11,24).
}
Both charges diverge as $k\rar\i$, and every theory lies strictly inside the conformal window, since
$\frac32<\frac{26}{17}<\frac{25}{14}<3$. Nevertheless, the exact changes in $a$ alternate in sign:
\e{}{
    \begin{aligned}
        a(B_k)-a(A_k)
        &=-\frac{327681}{6760000}\,4^k<0,\\
        a(A_{k+1})-a(B_k)
        &=\frac{789738849}{6760000}\,4^k>0.
    \end{aligned}
}
Meanwhile,
\e{}{
    a(A_k)=\frac{48657}{1250}\,4^k-\frac{3}{16},
    \qquad
    a(B_k)=\frac{420495}{10816}\,4^k-\frac{3}{16},
}
so $a\to\infty$ along the entire path. The negative increments persist at arbitrarily large charges and grow in magnitude. This gives an explicit CFT realization in which eventual $q$-ordering does not imply eventual central charge ordering.

\sec{The Katz-Sarnak conjecture for low-lying zeros of $L$-function families}\label{appe}

Writing the nontrivial zeros of a given $L(s,\pi_p)$ as $\rho_{\pi,j}=\tfrac12+it_{\pi,j}$, define an unfolded zero coordinate $x_{\pi,j}$ and its linear statistic as
\begin{equation}
x_{\pi,j}:=\frac{\log\mathfrak C(\pi)}{2\pi}t_{\pi,j},
\qquad
W_f(\pi):=\sum_j f(x_{\pi,j})
\label{eq:NT-zero-statistic}
\end{equation}
where $f$ is a real, even Schwartz test function. The unfolding probes zeros within order $1/\log\mathfrak C$ of the central point. For the Dirichlet family $\cF_{\rm Dir}$, $\mathfrak{C} \sim q$ as $q\gg1$. The so-called ``one-level density'' is the family mean of this linear statistic. At this level, the Katz--Sarnak prediction \cite{KS} takes the form
\begin{equation}
\mathbb E_N[W_f]
\longrightarrow
\int_{\mathbb R}f(x)\rho_{1,G}(x)\,dx
\label{eq:NT-one-level-density}
\end{equation}
as $N\rar\i$, where $G$ denotes the random-matrix symmetry type and $\rho_{1,G}$ its limiting one-point density. The full universality conjecture concerns the entire hierarchy of low-lying-zero statistics, not only the averaged linear statistic \eqref{eq:NT-one-level-density}. Schematically, for every fixed $n\geq1$ and suitable test function $F$ on $\mathbb R^n$, it predicts
\begin{equation}
\mathbb E_N\!\Bigg[
\sum_{j_1,\ldots,j_n}
F(x_{\pi,j_1},\ldots,x_{\pi,j_n})
\Bigg]
\longrightarrow
\int_{\mathbb R^n}F(\mathbf x)\,d\rho_{n,G}(\mathbf x)
\label{eq:NT-full-universality}
\end{equation}
as $N\rar\i$, where $\rho_{n,G}$ is the corresponding random-matrix correlation measure. 

In view of the CFT bound \eqr{v1}, which is the self-averaged version of the family gap bound \eqr{v5}, it is natural to ask if an $L$-function analog of the relation between the two has been established. Vis-\`a-vis Katz-Sarnak, the $L$-function analog would concern how far one can localize within the family $\cF_{\rm NT}$ while retaining a universal limiting law uniformly in increasingly narrow conductor windows. A concrete version of this was established by \cite{hughes} for the Dirichlet family $\cF_{\rm Dir}$, who considered the linear statistic $W_f$ defined in \eqr{eq:NT-zero-statistic} of the unfolded zeros of the $L$-functions of fixed prime $q$. What they showed was that, under some conditions, averaging over all $L$-functions at that value of $q$, then taking $q\rar\i$, yields moments that reproduce the unitary random matrix result.

\sec{Higher-spin gap}\label{appf}

Usually, $\Dgap$ is defined in an asymptotic large $N$ sense, where ``single-trace gap'' makes formal sense. As we will be working with paths out to large $N$, it will be important for us to take a more careful approach, in a manner that recovers the usual notion as $N\rar\i$.

Given a theory $\cT_N$, the Hilbert space of local primary operators (call it $\mathcal{H}_N$) may be graded by spin:
\e{}{\mathcal{H}_N = \bigoplus_\bJ \mathcal{H}_N^{(\bJ)}}
$\bJ$ is a multi-index, which indicates the presence of non-symmetric traceless irreducible representations of $SO(d)$ (relevant for $d>3$). Given a primary operator $\cO^{(\bJ)}$, define $J$ as the length of the first row of the Young diagram of the $SO(d)$ representation of $\cO^{(\bJ)}$. The meaning of ``higher-spin'' is $J>2$.\foot{One may also define $J$ as the exponent controlling the Regge limit of the conformal block of $\cO^{(\bJ)}$, scaling as $z^{1-J}$ as $z\rar 0$.}

As stated above, we would like to upgrade the usual treatment of $\Dgap$ to a notion that holds at sufficiently large $N$ -- in other words, to have an operational notion of ``approximately single-trace''. For a family with sufficiently regular large $N$ behavior of OPE data, this is intuitively simple, because the regularity allows us to ``track back'' from infinite $N$. However, we do not want to assume such regularity in general. 

The goal of a general definition of $\Dgap(N)$ is simple to state: in a given theory $\cT_N$, after ordering primaries by conformal dimension in every fixed spin sector, minimize over dimensions, then over spins $J>2$, subject to a non-compositeness constraint. The key is what constitutes non-compositeness. Some natural suggestions are too naive. For instance, suppose we order operators by dimension within every fixed $\bJ$ in every $\cT_N$, and construct branches $\mathscr{B}_i^{(\bJ)}$ that interpolate between operators of the $i$'th lowest dimension across the family: i.e. $\mathscr{B}_1^{(\bJ)}$ collects the lightest spin-$\bJ$ operators across all $N$, $\mathscr{B}_2^{(\bJ)}$ the second-lightest, and so on. We might then say to remove all branches whose $N\rar\i$ limit lands on a multi-trace higher-spin composite, and then to define $\Dgap(N)$ for $\cT_N$ as the lightest operator among those branches that remain. However, if a family $\cF$ has $\Dgap\rar\i$, then for any $\cT_N\in\cF$ of finite $N$, there could be {\it no} operators remaining at any finite $\D$: the rank $i$ of the branch to which the eventual $\Dgap$ is associated may grow with $N$. 

Instead, the goal is achieved by the following approach. With $N$ sufficiently large, the meaning of ``non-composite'' may be made quantitative, and operational, by enforcing a buffer $\eps_N$ to the dimensions of the would-be multiparticle Fock space at infinite $N$, where $\eps_{N\rar\i} \rar 0$. First, we define an effective gap in the spin-$\bJ$ sector as follows:
\e{effgapdef}{\Dgap^{(\bJ)}(N) := \inf\left\{\D_{\cO}(N) : \cO \in \mathcal{H}_N^{(\bJ)},~\left|\t_{\cO}(N) - \(2n+ \sum_i m_i \t_{\cO_i}(N)\)\right| \geq \eps_N~\,\forall\,\bigotimes_i \rho_{\cO_i}^{\otimes m_i} \supset \rho_\cO\right\}}
where $\t := \D-J$ is the twist, $m_i \in \Z_{\geq 0}, n\in \Z_{\geq 0}, \sum_i m_i \geq 2$, all operators $\cO_i \in \mathcal{H}_N$, and $\rho$ is a collective representation label for spin and any internal symmetries (and $\inf\varnothing=\i$). This non-compositeness condition is enforced on the spectral multiset, i.e. with multiplicity. Then
\e{}{\Dgap(N) := \inf_{J>2} \Dgap^{(\bJ)}(N)}
where we recall the definition of $J$ above. Taking large $N$ over a sequence defines the conventional higher-spin gap,
\e{}{\Dgap := \liminf_{N\rar\i} \Dgap(N)}
Here we have assumed that the minimizing spin stabilizes (if not, swap the inf and $\liminf$); in typical large $N$ examples, $J\leq 4$. The precise choice of $\eps_N$ is not unique. In a $1/c$ expansion, anomalous dimensions of multi-trace operators admit a power series in $1/c$, with leading-order scaling $\g_{n,J} \approx f(n)/c$. This motivates the choice $\eps_N = K/ c_N$ for some $N$-independent $K<\i$, though further refinement is possible.\foot{There is one technical issue which this construction has not yet addressed, which is that $f(n)$ grows with $n$: from holography, it is known that $f(n) \sim n^{d-1}$ as $n\gg 1$ \c{Cornalba:2006xm,MSZ}. A rigorous recursive procedure that overcomes this, properly removing the entire tower of composite states, even those for which $n \gg 1$, is as follows. First, we identify a ``seed'' composite, of some twist $\t_0$ that fails the inequality in \eqr{effgapdef} for $n=0$; next, we initialize a sequence of primaries, of twists $\t_0,\t_1,\ldots$, that are subject to successive finite-difference equations with the same error $\eps_N$. Let us demonstrate with an example. Suppose a double-trace operator $\cO$ has $\tau_n \approx 2\t_{\cO'} + 2n+{n^2/ c}$. When $n\sim \sqrt{c}$, this would exceed the tolerance in \eqr{effgapdef}, despite being a genuine double-trace state. On the other hand, the three-term relation obeys $|\tau_{n+1}-2\tau_n+\tau_{n-1}| \approx {2/c}$: thus, a compositeness test based on enforcing the same tolerance $\eps_N$ on this finite difference would correctly reject this double-trace state.} Note that in the presence of commensurable conformal weights, the prescription should account for ``accidental'' degeneracies between composite and non-composite large $N$ operators. Relatedly, for slightly broken higher-spin CFTs, such as the ungauged critical $O(N)$ model in $d=3$, the definition should identify $\Dgap$ with the dimension of the $J=4$ almost-conserved current, $\D_{J=4} \approx 5 + O(1/N)$.

\sec{Further comment on families and central charges}\label{appg}

Why must we talk about families in order to take large central charge? 

One answer, independent of holography, is the $c$-theorem, which implies that $c$ cannot vary continuously as a function of parameters: a ``single CFT'' has a single value of $c$. This is rigorous in $d=2$ (with some caveats), and one could reasonably quibble about whether, and for precisely which quantity, it is true in general dimension $d$. However, considering the totality of available evidence, a robust version of this statement seems to hold, no matter which quantity we identify as ``$c$''. 

In $d=2$, Zamolodchikov's proof of this statement assumes that the putative parameters are exactly marginal couplings. It is an open problem to prove that in general 2d CFT, $c$ cannot depend on any continuous parameters at all. This is an interesting problem from a bootstrap point of view because, in view of ``counterexamples'' like Liouville CFT, it is not clear precisely which abstract conditions should be responsible for the apparent independence of $c$ of continuous parameters.\foot{A reasonable conjecture is that the statement is true for all unitary, compact 2d CFTs. One can make a fairly sharp principled argument to this effect (available upon request) upon further assuming random matrix statistics at high energy.} A different approach would be to prove that {\it all} continuous parameters must be exactly marginal couplings, or at least under certain conditions; see \c{KomatsuKusukiMeineriOoguri} for recent progress in this direction in a context assuming certain conformal interfaces.

In $d>2$, there is no proof that $C_T$ cannot vary along a conformal manifold. But there is also no example of a unitary, local $d>2$ CFT in which $C_T$ has been explicitly certified to vary continuously. This brings up the usual point that non-supersymmetric conformal manifolds make this more plausible (e.g. in $d=4$, supersymmetry does enforce $\partial_\lambda c=0$), but are harder to construct. Even if $C_T$ could vary, one would not expect {\it unbounded} variation to be possible, in the spirit of the Hofman-Maldacena collider bounds \c{HM}. Perhaps the situation is different in odd dimensions. 

In any case, the need for families is clear for understanding AdS quantum gravity, in which Newton's constant is quantized. 

\end{appendix}

\end{spacing}

\bibliographystyle{JHEP}
\bibliography{N_refs.bib}

\end{document}